\documentclass[fleqn,usenatbib]{mnras}

\usepackage{newtxtext,newtxmath}

\usepackage[T1]{fontenc}

\DeclareRobustCommand{\VAN}[3]{#2}
\let\VANthebibliography\thebibliography
\def\thebibliography{\DeclareRobustCommand{\VAN}[3]{##3}\VANthebibliography}

\usepackage{makecell}
\usepackage{caption}
\usepackage{subcaption}
\usepackage{hyperref}
\usepackage{float}
\usepackage{adjustbox}

\usepackage{graphicx}	
\usepackage{amsmath}	
\usepackage{braket}
\usepackage{xcolor}
\usepackage{gensymb}
\usepackage{hyperref}
\hypersetup{
    colorlinks=true,
    linkcolor=blue,
    filecolor=magenta,      
    urlcolor=cyan,
    pdftitle={Overleaf Example},
    pdfpagemode=FullScreen,
    }

\newcommand\Msol{$M_\odot$}

\title[Magnetic Fields in Bok Globules]{A Survey of Magnetic Field Properties in Bok Globules}

\author[Roychowdhury et al.]{Tamojeet Roychowdhury$^{1}$\thanks{E-mail: tamojeet@iitb.ac.in},
Thushara G.S. Pillai$^{2}$,
Claudia Vilega-Rodrigues$^3$,
Jens Kauffmann$^2$,
\newauthor{Le Ngoc Tram$^4$,
Tyler L. Bourke$^5$
and Victor de Souza Magalhaes$^6$}
\\ \\
$^{1}$Department of Electrical Engineering, Indian Institute of Technology Bombay, Mumbai 400076 India\\
$^{2}$Haystack Observatory, Massachusetts Institute of Technology, 99 Millstone Road, Westford, MA~01886, USA\\
$^{3}$ Instituto Nacional de Pesquisas Espaciais (INPE/MCTI), Av. dos Astronautas, 1758, S\~ao Jos\'e dos Campos, SP, Brazil\\
$^4$ Leiden Observatory, Leiden University, PO Box 9513, 2300 RA Leiden, The Netherlands\\
$^5$ SKA Observatory, Jodrell Bank, Lower Withington, Macclesfield, Cheshire, SK11 9FT, UK \\
$^6$ National Radio Astronomy Observatory, 800 Bradbury Dr., SE Ste 235, Albuquerque, NM 87106, USA
}

\pubyear{2024}

\begin{document}
\label{firstpage}
\pagerange{\pageref{firstpage}--\pageref{lastpage}}
\maketitle

\begin{abstract}
Bok globules are small, dense clouds that act as isolated precursors for the formation of single or binary stars. Although recent dust polarization surveys, primarily with   \textit{Planck}, have shown that molecular clouds are strongly magnetized, the significance of magnetic fields in Bok globules has largely been limited to individual case studies, lacking a broader statistical understanding. In this work, we introduce a comprehensive optical polarimetric survey of 21 Bok globules. Using \textit{Gaia} and near-IR photometric data, we produce extinction maps for each target. Using the radiative torque alignment model customized to the physical properties of the Bok globule, we characterize the polarization efficiency of one representative globule as a function of its visual extinction. We thus find our optical polarimetric data to be a good probe of the globule's magnetic field. Our statistical analysis of the orientation of elongated extinction structures relative to the plane-of-sky magnetic field orientations shows they do not align strictly parallel or perpendicular. Instead, the data is best explained by a bimodal distribution, with structures oriented at projected angles that are either parallel or perpendicular. The plane-of-sky magnetic field strengths on the scales probed by optical polarimetric data are measured using the Davis-Chandrasekhar-Fermi technique. We then derive magnetic properties such as Alfvén Mach numbers and mass-to-magnetic flux ratios. Our findings statistically place the large-scale (A$_{\mathrm{V}} < 7 \, \text{mag}$) magnetic properties of Bok globules in a dynamically important domain.
\end{abstract}




\section{Introduction}

Bok globules, named after the astronomer Bart Bok \citep{Bok_1947, Bok_1948}, are small, isolated dark clouds that play a role in low-mass star formation. They have representative masses of $\sim$1–10~\Msol\ within $<1$\,pc and hosts only one or two dense cores \citep{Kandori_2005, Reipurth_2008, Launhardt_2010}. Compared to giant molecular clouds with multiple sites of star formation and considerable feedback, Bok globules therefore serve as simpler environments to constrain the physics of star formation. Multi-wavelength studies have revealed thermal emission from dust \citep{Clemens_1988, Clemens_1991, BHR_1995, Moreira_1997, Moreira_1999, Launhardt_Henning_1997, Henning_Launhardt_1998}, submillimeter emission from protostars \citep{Huard_1999, Sadavoy_2018, Pattle_2022}, and near-infrared observations identifying young stellar objects (YSOs) \citep{Yun_1995, Alves_1994, Racca_2009} in some Bok Globules. Molecular line observations provide insights into physical conditions, identifying collapsing clouds and molecular outflows \citep{Wang_1995, Yun_1992, Yun_1994, Marka_2012}. 

The study of magnetic field strength in Bok globules is crucial for understanding the star formation process. Dust grains in the interstellar medium (ISM) are responsible for the polarization of starlight \citep{Hall_1949, Hiltner_1949} and polarized thermal emission (e.g. \citealt{Planck_2016}) caused by such aligned dust grains is the most widely used method for probing the projected magnetic fields on the plane of the sky (POS) \citep{Andersson_2015, Tram_2022}. A growing body of theoretical and observational evidence now supports the radiative alignment torque (RAT) mechanism as the most likely explanation for the observed optical/near-infrared (O/IR) and far-infrared (FIR)/sub-millimeter (submm) polarization in the ISM \citep{Draine_1997, Lazarian_2007}, wherein magnetic fields align non-spherical dust grains through radiative torques. O/IR starlight polarimetry efficiently probes  the large scale magnetic field structure of Bok Globules \citep{Jones_1984, Klebe_1990, Kane_1995, Sen_2000, Sen_2005, Bertrang_2014, Das_2016, Kandori_2020}. FIR/submm polarization maps have revealed magnetic field distribution towards dense cores within Bok Globules \citep{Davis_2000, Henning_2001, Vallee_2003, Wolf_2003, Ward-Thompson_2009, Zielinski_2021, Yen_2020, Pattle_2022}. These polarization patterns display significant diversity with respect to cloud structure, with the degree of polarization decreasing toward the dense cores. 
Such studies also applied the Davis-Chandrasekhar-Fermi (DCF; \citealt{Davis_1951, Chandrasekhar_Fermi_1953}) formalism to estimate magnetic field strengths in the range of a few hundred $\mu$G \citep{Davis_2000, Henning_2001, Vallee_2003, Wolf_2003, Zielinski_2021},  suggesting a important role of magnetic fields in these environments.  Polarization studies conducted so far have been  largely case studies or involving at most a handful of objects. A dedicated polarization survey that can explore trends across various physical properties of Bok Globules is critically missing.

This paper is based on optical polarization data from the same sample of 21 Bok Globules analyzed by \citet{Racca_2009}. A preliminary analysis of these data shows a tendency of globules having young stellar objects have less organized magnetic fields than the quiescent ones \citep{2014IAUS..302...21R}.
These globules are not associated with bright nebulae or molecular complexes.
By systematically analyzing for the first time a large sample of starlight polarimetric maps of Bok globules, this study seeks to address several unresolved questions: To what extent do magnetic fields dictate the morphology of Bok globules?  Are there systematic differences in magnetic field structure and strength across different evolutionary stages of Bok Globules with and without star formation? These questions are explored by combining \textit{Gaia} and   \textit{Planck} based dust column density data and optical starlight polarization data, allowing for a more detailed understanding of the interplay between magnetic fields and star formation in Bok globules.

This paper is organized as follows. In Section~\ref{sec:obs}, we outline the observational data utilized in this study. Section~\ref{sec:methods} details the methodologies adopted, including the extraction of cloud orientations (Section~\ref{sec:rht}), the generation of extinction maps (Section~\ref{sec:meth_av}), the modeling of dust grain alignment efficiency (Section~\ref{PE-ext}), and the derivation of magnetic field strengths (Section~\ref{sec:meth_B}). Our findings based on these methods are presented in Section~\ref{sec:results}, followed by an interpretation of the results in Section~\ref{sec:disc}. Finally, we summarize the key conclusions of this work in Section~\ref{sec:conc}.

\section{Observations }\label{sec:obs}

We have obtained optical polarimetric data of a sample of Southern Bok Globules at the Observatório do Pico dos Dias (OPD), using the 0.6-m Boller \& Chivens telescope and the IAGPOL polarimeter \citep{Magalhaes1996}, which was equipped with a rotating half-wave retarder plate. The CCD used was an Ikon~L, with 2048 $\times$ 2048 pixels, configured with a 1~MHz readout frequency and a pre-amplifier gain of 4. This reading mode results in a gain of 0.9~electrons per ADU and a readout noise of 6~electrons. The data were collected using a $I_c$ filter (centered at around 840~nm with a FWHM of 150~nm) to minimize the effect of the interstellar extinction. The images have a field of view of 11~$\times$~11~arcmin$^2$ without the focal reducer and twice this value with the reducer. The observations log is shown in Table~\ref{tab_obs}. Polarimetric standard stars were also observed to correct the instrumental polarization angle to the equatorial system.

The used polarimetric technique splits the incident light in two beams of orthogonal polarizations, producing the so called ordinary and extraordinary images of a given source. The ratio between the difference and the sum of the counts of those images depends on the polarization of the source and also on instrumental configuration. Specifically, this quantity is a function of the $Q$ and $U$ Stokes parameters and modulates as a function of the retarder position. Therefore, the observed modulation can be used to determine the source polarization. This technique naturally removes the sky polarization from the estimated Stokes parameters of the source. The interested reader can see more on this dual-beam polarimetric technique in \citet{Magalhaes1996} and \citet{1984PASP...96..383M}. 

The reduction was performed using standard IRAF routines \citep{iraf1,iraf2} to perform bias and dome flat-fields corrections as well as aperture photometry of the ordinary and extraordinary images. The polarization estimates were performed using the {\sc pccdpack} \citep{pereyra2000,pereyra2018} and {\sc pcckpack\textunderscore inpe}\footnote{\url{https://github.com/claudiavr/pccdpack_inpe}} IRAF packages.  The polarimetric data was originally presented in \citet{magalhaes2012}, where the reduction process is described in detail.

\begin{table}
\begin{adjustbox}{width=1.1\columnwidth,center}
\begin{tabular}{|c|c|c|c|c|c|}

\hline
Object & Date   & Exposure   & Number of & Focal & Distance (pc) \\
       &        &  Time (s)  & exposures & reducer & \\
\hline
BHR~016 & 2011 May 03 & 40 & 12 & N & 300\\
BHR~034 & 2011 May 03 & 40 & 12 & N & 400\\
BHR~044 & 2011 May 03 & 40 & 12 & N & 300\\
BHR~053 & 2010 Jun 01 & 30 & 8 & N & 500\\
BHR~058 & 2010 Jun 01 & 200 & 8 & N & 250\\
BHR~059 & 2010 Jun 01 & 40 & 16 & N & 250\\
BHR~074 & 2010 May 31 & 30 & 8 & Y & 175\\
        & 2010 May 31 & 180 & 8 & Y & \\
BHR~075 & 2010 May 31 & 30 & 8 & Y & 175\\
        & 2010 May 31 & 200 & 8 & Y & \\
BHR~111 & 2010 Jun 01 & 200 & 8 & N & 250\\
BHR~113 & 2010 Jun 01 & 200 & 8 & N & 200\\
BHR~117 & 2010 Jun 01 & 200 & 8 & N & 250\\
BHR~121 & 2011 May 03 & 30 & 12 & N & 300\\
BHR~126 & 2011 May 03 & 40 & 16 & N & 170\\
BHR~133 & 2010 Jun 01 & 200 & 8 & N & 700\\
BHR~138 & 2011 May 05 & 40 & 12 & N & 400\\
BHR~139 & 2011 May 05 & 40 & 12 & N & 400\\
BHR~140 & 2011 May 05 & 40 & 12 & N & 400\\
BHR~144 & 2011 May 03 & 40 & 12 & N & 170\\
BHR~145 & 2011 May 05 & 40 & 12 & N & 450\\
BHR~148 & 2011 May 03 & 40 & 16 & N & 200\\
BHR~149 & 2011 May 04 & 40 & 12 & N & 200\\
\hline
\end{tabular}
\end{adjustbox}
\caption{Log of observations for the polarization observations and their respective distances.}
\label{tab_obs}
\end{table}  

\section{Methods} \label{sec:methods}

We primarily need two physical quantities for each target -- the filament structure (to extract relative orientation w.r.t. the ambient magnetic field) and the average extinctions (that gives us an estimate of the density and mass in the region).

For structures, the ideal resolution and sensitivity is provided by Herschel, but due to unavailability of these for all of our 21 targets, we use \textit{Gaia} stellar density maps in the region (which also correlates well with the \textit{Herschel} structures, where available). 

For extinctions, we can estimate them from near-infrared reddening with 2MASS, or use emission-modeling derived values with   \textit{Planck}.   \textit{Planck}, however, has a low resolution and is also contaminated by background galactic emission. We thus use 2MASS-derived values for our targets, also finding a good agreement with   \textit{Planck}-derived values where computable.

The procedures of creating maps and calculating these quantities is detailed in the following subsections.

\subsection{Structure Determination }
\label{sec:rht}

\textit{Herschel} maps of infrared dust emission at wavelengths of 250, 350 and 500 $\mu$m from the SPIRE instrument \citep{Griffin_2010} provided ideal images with fine spatial resolution to identify the dominant structure of each Bok globule. However, \textit{Herschel} data were available only for six of our 21 globules: BHR 16, 34, 53, 74, 111 and 140. Since this would reduce our sample size drastically, we opted to use an alternative technique based on star counts to ascertain the cloud structure.

\begin{figure}
	\includegraphics[width=0.47\columnwidth]{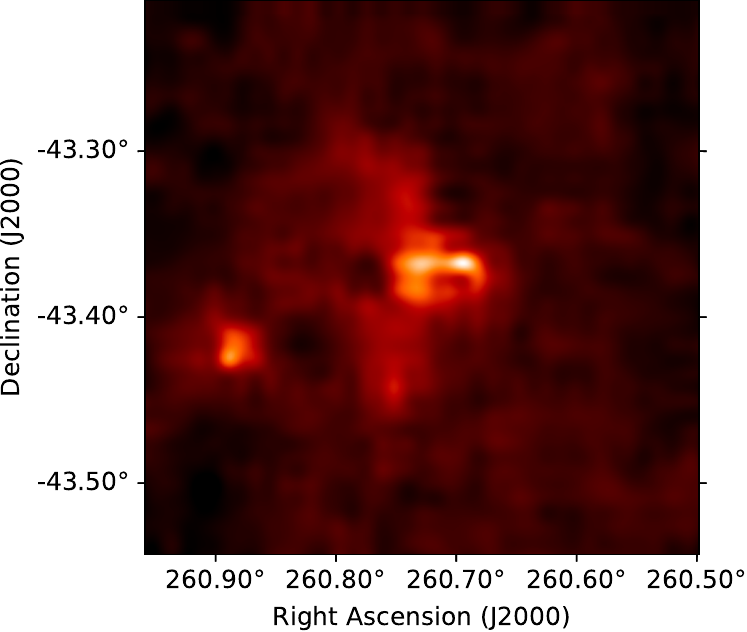} 
    \includegraphics[width=0.47\columnwidth]{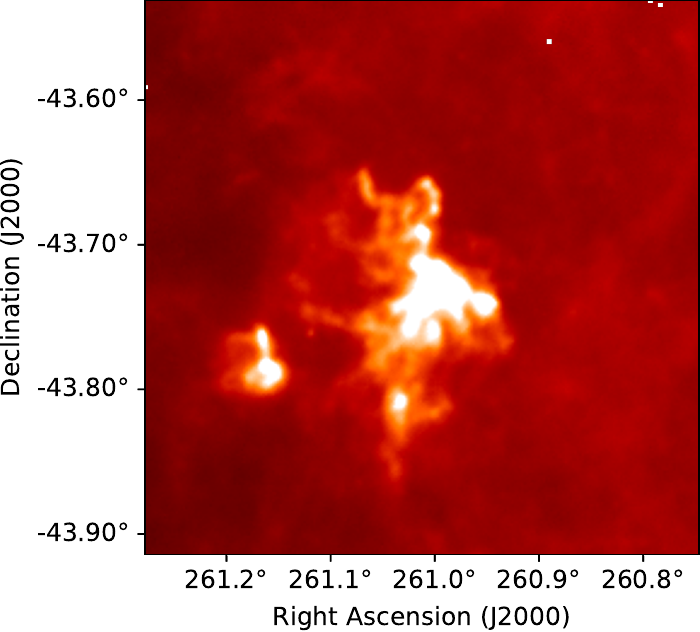} \\
    \includegraphics[width=0.47\columnwidth]{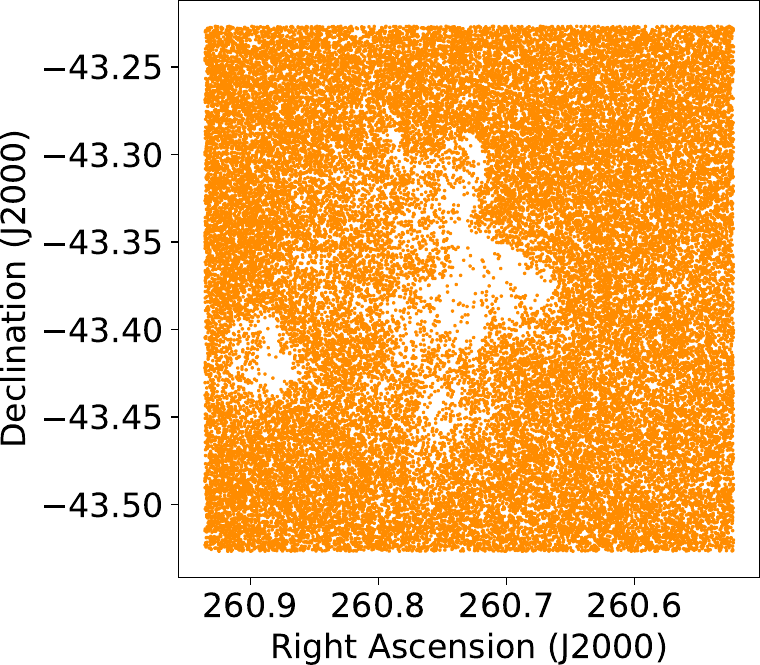} 
    \includegraphics[width=0.47\columnwidth]{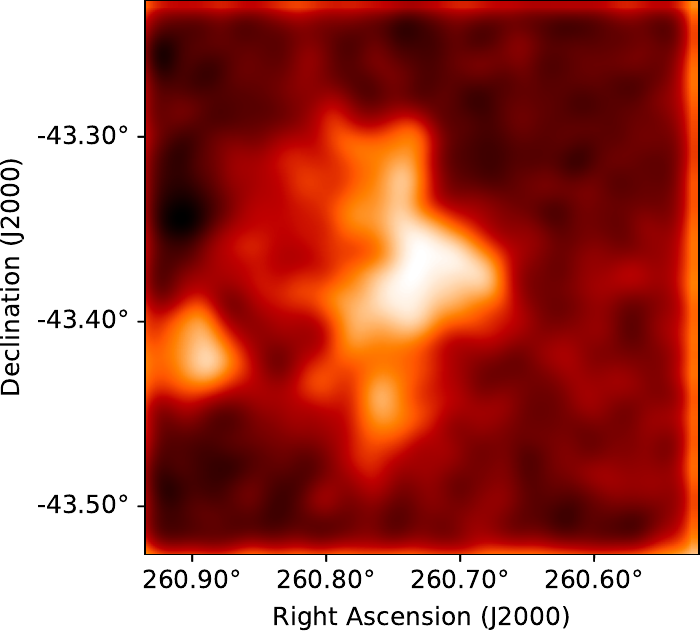}
    \caption{Comparison between maps of BHR 140 obtained from various telescopes. Top left - 2MASS $(H-K)$; Top right - \textit{Herschel} 250 $\mu$m, Bottom left - raw \textit{Gaia} star density; Bottom right - smoothed (binned and convolved) \textit{Gaia} star density.}
    \label{fig:map_comparisons}
\end{figure}

We used the optical catalog of stars from \textit{Gaia} DR3 and queried around each Bok globule within a square of size 18 \arcmin $\times{}$18 \arcmin centered on the location of the globule obtained from the BHR catalogue \citep{BHR_1995}. The distances to the globules were taken from the same reference. We found almost no stars present along the line-of-sight closer than the distance (inverse of \textit{Gaia} parallax) of the globule itself. We also noticed that each of the globules appears as a 'hole' devoid of any stars in optical visibility, in the RA-Dec space (see bottom-left panel of Fig~\ref{fig:map_comparisons} for an example). Thus we attempted to define the boundary of the large scale structure of the globule using the extent of this 'hole' following the procedure outlined below:

\begin{itemize}
    \item All stars within a square of size $0.3\times{}0.3{\ }\mathrm{deg}^2$ centred at the globule were queried for from \textit{Gaia} DR3.
    \item The corresponding RA-Dec space was divided into a $60\times60$ grid of 'pixels', each pixel thus measuring $18\arcsec$. 
    \item Each pixel was assigned a number equal to the number of stars lying within that pixel. The hole that defined the Bok globule outline had all pixels with values equal to 0. The pixel values were then inverted, i.e. the pixels with zero stars were assigned the maximum value of the 2D array and vice-versa.
    \item To reduce noise, the pixel values were smoothed by convolving with a 2D Gaussian kernel with $\sigma=1.5~\rm{}pixels$.
    \item Finally, the RA-Dec grid was sectioned into a finer grid of $300\times{}300$, each new pixel now being $3\farcs{}6$ in side, and intermediate values were assigned by a 2D cubic spline interpolation
\end{itemize}

This method worked well enough to return not only the large scale structure of each cloud, but also several surrounding auxiliary filament-like structures of lower column densities with an accuracy comparable to the 2MASS maps in \cite{Racca_2009}. We do note, however, that these \textit{Gaia} stellar density maps in no way can be a proxy for visual extinction maps, and shall only be used for defining cloud structure and boundary. These maps were compared against the \textit{Herschel} maps for the six clouds where SPIRE data were available, and had an excellent match for both the main cloud as well as the less dense filament structures around it as revealed by visual inspection (see, for instance, Fig~\ref{fig:map_comparisons} for BHR 140). These are hence useful to demarcate the low and high extinction regions qualitatively.

We then tried to define the cloud's structure (by broadly taking each cloud to have an elliptical projection on the plane of sky) using the major and minor axis and its position angle (PA). The \textit{Gaia} maps have a finer resolution where curved structures cannot be fitted easily to an ellipse, so we needed to reduce the resolution to get an elliptical shape. We convolved our original \textit{Gaia} maps with a Gaussian kernel of $\sigma=0.08~\rm{}pc$ size, with the angular pixel scale adjusted using \cite{Racca_2009} distances.

\begin{figure*}
	\includegraphics[width=0.24\linewidth]{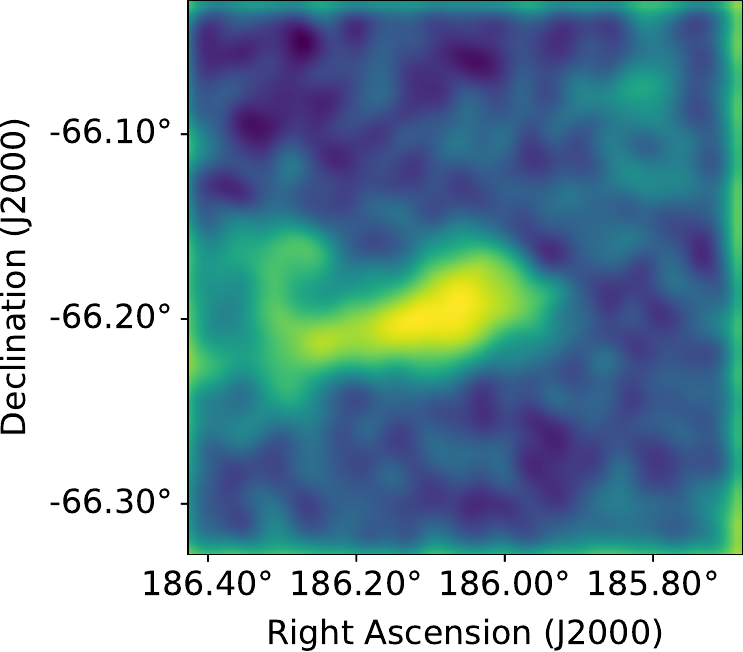}
    \includegraphics[width=0.24\linewidth]{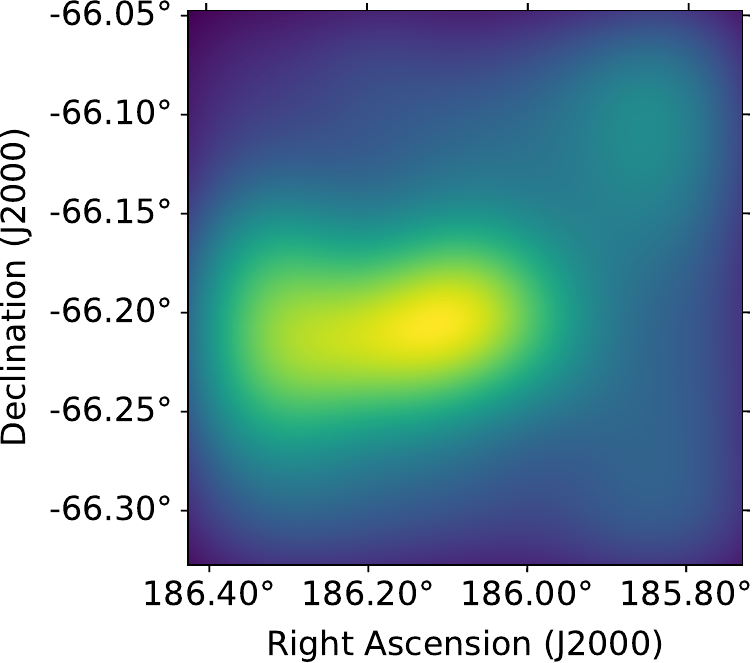}
    \includegraphics[width=0.24\linewidth]{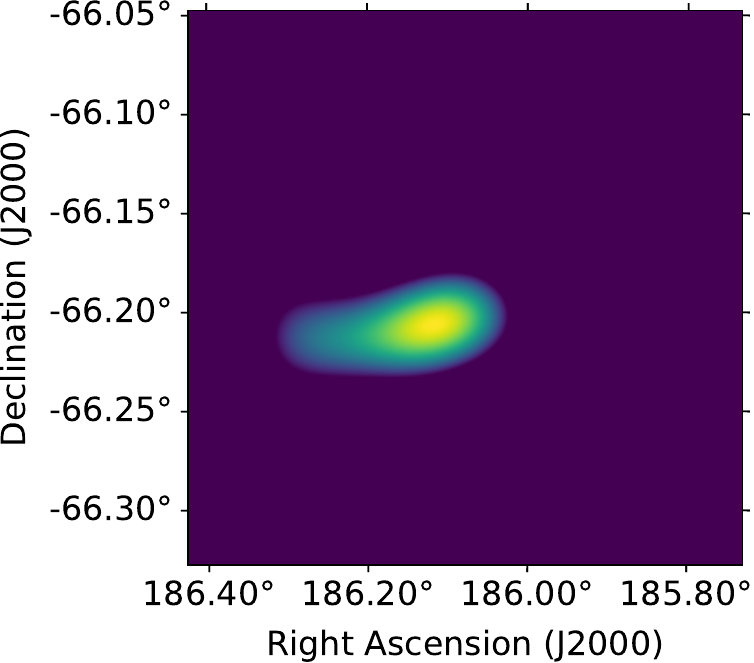}
    \includegraphics[width=0.24\linewidth]{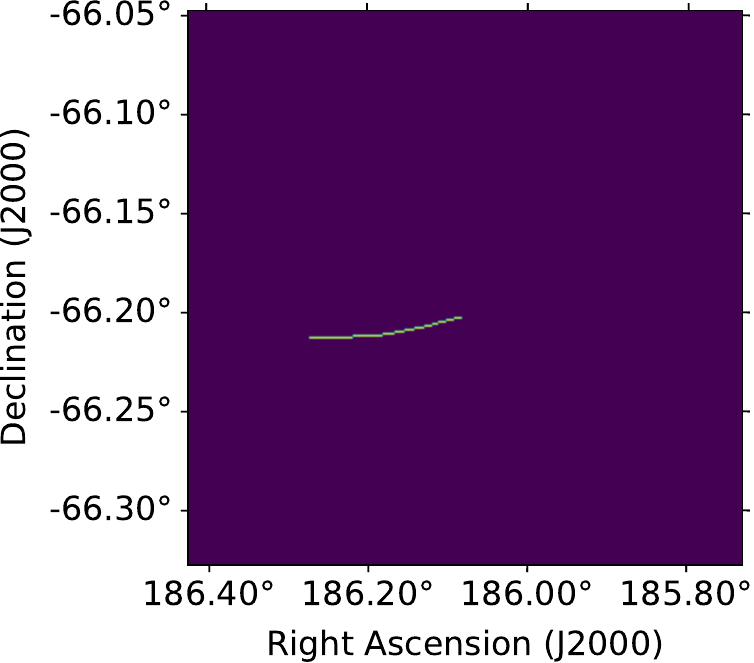}
    \caption{An outline of the procedure used to find the dominant orientation for each cloud upto filament extraction (for BHR 75 in this figure, images from left to right): creating \textit{Gaia} stellar density maps, convolving it to smooth out boundaries into a roughly elliptical shape, masking at a threshold to extract the ellipse, running the \texttt{fil-finder} Python algorithm}
    \label{fig:gaia-structure}
\end{figure*}

These low resolution maps were then fed into the \texttt{FilFinder} algorithm on Python, designed by \cite{Koch_2015}, which effectively isolated the dominant structure as a single filament. This procedure is depicted in Fig~\ref{fig:gaia-structure} To find the position angle, we applied the Rolling Hough Transform (RHT) as described in \cite{Clark_2014} on the filaments returned by \texttt{FilFinder}. The RHT algorithm gives the power of each possible angle along the filament, described in detail at \url{https://seclark.github.io/RHT/}. A power-weighted mean PA of all points along the entire filament structure was taken to define the cloud's PA.

Once the PA is obtained, we found the major and minor axis of the fitting ellipse as follows: the interior (and hence the boundary) of the ellipse was defined to be encompassing all pixels whose value exceeded a certain threshold. In most cases this threshold was set as 0.88-0.94 times the maximum pixel value. Two clouds (BHR 148 and 149) had slightly different thresholds owing to background substructures unrelated to the actual globule. The major axis is then calculated as the length of a straight line of pixels through the centre of the cloud, at the angle given by RHT. The minor axis is found the same way at an angular perpendicular to the PA. 

Figure~\ref{fig:gaia-structure} illustrates the steps of the procedure described above. The obtained position angles for the primary filament are listed in Table~\ref{tab:ellipse-fits}.

\begin{table}\centering

\begin{tabular}{|c|c|c|c|} \hline
Name & Position Angle (\degree) & Major axis (pc) & Minor axis (pc) \\ \hline
BHR 16  &  76.5 $\pm$ 16 & 0.51 & 0.3 \\
BHR 34  &  115.19 $\pm$ 27 & 0.34 & 0.25 \\
BHR 44  &  83.6 $\pm$ 10 & 0.67 & 0.46 \\
BHR 53  &  13.52 $\pm$ 27 & 0.45 & 0.26 \\
BHR 58  &  57.74 $\pm$ 14 & 0.21 & 0.15 \\
BHR 59  &  68.24 $\pm$ 5 & 0.27 & 0.24 \\
BHR 74  &  127.4 $\pm$ 21 & 0.17 & 0.15 \\
BHR 75  &  104.81 $\pm$ 16 & 0.37 & 0.15 \\
BHR 111  &  129.85 $\pm$ 5 & 0.71 & 0.41 \\
BHR 113  &  136.96 $\pm$ 4 & 0.15 & 0.1 \\
BHR 117  &  42.04 $\pm$ 20 & 0.53 & 0.3 \\
BHR 121  &  78.0 $\pm$ 11 & 0.57 & 0.15 \\
BHR 126  &  98.4 $\pm$ 15 & 0.18 & 0.17 \\
BHR 133  &  61.15 $\pm$ 12 & 1.3 & 0.51 \\
BHR 138  &  6.34 $\pm$ 17 & 0.22 & 0.2 \\
BHR 139  &  55.98 $\pm$ 15 & 0.24 & 0.22 \\
BHR 140  &  21.68 $\pm$ 12 & 0.61 & 0.43 \\
BHR 144  &  128.54 $\pm$ 20 & 0.31 & 0.2 \\
BHR 145  &  99.5 $\pm$ 9 & 0.63 & 0.43 \\
BHR 148  &  82.83 $\pm$ 22 & 0.22 & 0.18 \\
BHR 149  &  43.62 $\pm$ 7 & 0.17 & 0.09 \\

\hline
\end{tabular}

\caption{Properties of the best fit ellipse for each Bok globule. Position angle is measured East of North}\label{tab:ellipse-fits}
\end{table}

\subsection{Extinction Maps and Mass Determination} \label{sec:meth_av}

\begin{figure}
	\includegraphics[width=\columnwidth]{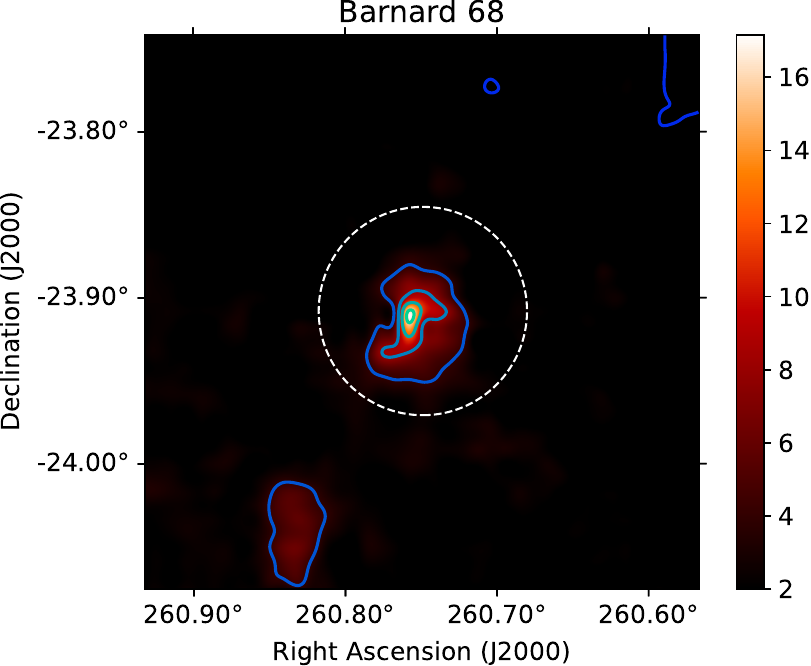}
    \caption{Application of our mass determination method to the Barnard 68 Bok globule. Colorbar denotes the visual extinction, and circular region marks the area used for mass estimation}
    \label{fig:2mass-ext}
\end{figure}

We use a simplified version of the method outlined in \cite{Racca_2009} using 2MASS magnitudes to create extinction maps. Having high spatial resolution was not a primary objective since structure is determined well with \textit{Gaia} as described above. Our procedure went as follows.

\begin{itemize}
    \item All stars lying within a $20\arcmin\times~20\arcmin$ square centred at the globule's coordinates are queried from 2MASS.
    \item The RA-Dec space is divided into a grid of pixels of $0\farcm{}5$ size. This, however, left several pixels empty, so we could not create a map stars only within each individual pixel.
    \item For the same grid with RA/Dec spacing of $0\farcm{}5$, we select all stars in a $1\farcm{}5$ square of each grid point. This leads to a repetition in the star sample between neighbouring grid points, but does not return an empty set for any grid point. The average colours of these stars, $\overline{H-K}$, where $H$ and $K$ are the standard near-infrared 2MASS bands at 1.65 and $2.19~\mu\rm{}m$, was taken as the representative value of reddening.
    \item The average intrinsic colour of a star, taken as $0.13$ from \cite{Lada1994}, was subtracted from this $\overline{H-K}$. The average reddening is thus $\overline{H-K}-0.13$.
    \item The larger area for selecting stars ($1\farcm{}5$) ensured a smooth map so separate Gaussian smoothing was unnecessary. However, the pixels were more finely divided into $3\farcs{}6$ each, with cubic interpolation, similar to our procedure for \textit{Gaia} maps.
    \item Each pixel number corresponds to $E(H-K)=\overline{H-K}-0.13$, where $E(.)$ denotes the excess due to reddening. We convert it to visual extinction as $A_V = 15.9\times(\overline{H-K}-0.13)$, as given in \cite{Lada1994}.
\end{itemize}

\cite{Racca_2009} however noted that the minimum measurable visual extinction $A_V$ varied from 2 to 4 depending on the error in $(H-K)$ measurements. We adopt a value of 3 -- all pixels with $A_V > 3$ and lying within a circle of a specific radius around the centre represent the Bok globule. The centre coordinates and the radii were tweaked slightly for different globules to accommodate for their sizes and orientations.

To translate the visual extinction to column density and subsequently to masses, we use the relation 
$$ \frac{N_\mathrm{H_2}}{A_V} = 9.4 \times 10^{20} \  \mathrm{mag}^{-1}$$
as given by \cite{Bohlin_1978}. This gives us the column density for each pixel. Distances from \cite{Racca_2009} are used to convert the angular scale to length scale for each cloud, which subsequently gives us the number density of H$_2$ as $n_\mathrm{H_2} = N_\mathrm{H_2}/l$, where $l$ is taken as the projection of $10\arcmin$ at the cloud distance, since our polarization measurements probe a region of $10\arcmin\times10\arcmin$ around the cloud. The mass density is then calculated as $\rho=1.36\,m_{\rm{}H_2}\,n_{\rm{}H_2}$, where $m_{\rm{}H_2}\approx{}2~{\rm{}amu}$ and the factor 1.36 is to account for the distribution of heavier elements including Helium \citep{Kauffmann_2008}. Mass is summed over all pixels with $A_V > 3$ and inside the circle.

\begin{table}\centering
\begin{adjustbox}{width=1.1\columnwidth,center}
\begin{tabular}{|c|c|c|c|} \hline
Name & Column Density $N_\mathrm{H_2}$ & Number Density $n_\mathrm{H_2}$ & Mass Density $\rho$ \\
 & in $10^{20}$ H$_2$ cm$^{-2}$ & in $10^{3}$ H$_2$ cm$^{-3}$ & in $10^{-21}$ g cm$^{-3}$ \\ \hline
BHR 16  &  19.5 $\pm$ 9.7 & 0.7 $\pm$ 0.4 & 3.3 $\pm$ 1.9 \\
BHR 34  &  12.9 $\pm$ 6.5 & 0.4 $\pm$ 0.1 & 1.6 $\pm$ 0.4 \\
BHR 44  &  22.3 $\pm$ 11.2 & 0.8 $\pm$ 0.3 & 3.7 $\pm$ 1.3 \\
BHR 53  &  16.1 $\pm$ 8.1 & 0.4 $\pm$ 0.1 & 1.6 $\pm$ 0.6 \\
BHR 58  &  4.8 $\pm$ 2.4 & 0.2 $\pm$ 0.1 & 1.0 $\pm$ 0.5 \\
BHR 59  &  23.9 $\pm$ 12.0 & 1.1 $\pm$ 0.5 & 4.8 $\pm$ 2.2 \\
BHR 74  &  5.4 $\pm$ 2.7 & 0.3 $\pm$ 0.2 & 1.6 $\pm$ 0.9 \\
BHR 75  &  6.6 $\pm$ 3.3 & 0.4 $\pm$ 0.2 & 1.9 $\pm$ 0.9 \\
BHR 111  &  39.5 $\pm$ 19.8 & 1.8 $\pm$ 0.5 & 8.0 $\pm$ 2.2 \\
BHR 113  &  51.5 $\pm$ 25.8 & 2.9 $\pm$ 2.7 & 13.0 $\pm$ 12.0 \\
BHR 117  &  12.1 $\pm$ 6.0 & 0.5 $\pm$ 0.2 & 2.4 $\pm$ 0.8 \\
BHR 121  &  1.6 $\pm$ 0.8 & 0.1 $\pm$ 0.1 & 0.3 $\pm$ 0.3 \\
BHR 126  &  17.7 $\pm$ 8.9 & 1.2 $\pm$ 0.5 & 5.3 $\pm$ 2.4 \\
BHR 133  &  60.0 $\pm$ 30.0 & 1.0 $\pm$ 0.4 & 4.3 $\pm$ 1.9 \\
BHR 138  &  13.0 $\pm$ 6.5 & 0.4 $\pm$ 0.2 & 1.6 $\pm$ 0.7 \\
BHR 139  &  12.6 $\pm$ 6.3 & 0.3 $\pm$ 0.1 & 1.6 $\pm$ 0.4 \\
BHR 140  &  10.7 $\pm$ 5.3 & 0.3 $\pm$ 0.1 & 1.3 $\pm$ 0.5 \\
BHR 144  &  11.3 $\pm$ 5.7 & 0.7 $\pm$ 0.5 & 3.4 $\pm$ 2.4 \\
BHR 145  &  10.1 $\pm$ 5.0 & 0.3 $\pm$ 0.1 & 1.1 $\pm$ 0.5 \\
BHR 148  &  27.2 $\pm$ 13.6 & 1.5 $\pm$ 0.3 & 6.8 $\pm$ 1.4 \\
BHR 149  &  24.6 $\pm$ 12.3 & 1.4 $\pm$ 0.6 & 6.2 $\pm$ 2.9 \\

\hline
\end{tabular}
\end{adjustbox}
\caption{Density of the Bok globules calculated from 2MASS extinctions}\label{tab:densities}
\end{table}

To test the accuracy of this method, it was first applied on Barnard 68, a nearby and well-studied Bok globule, whose distance and mass are known to be 125 pc and about $2\,M_\odot$ respectively, from \cite{Alves_2001}. Our method with the above adopted distance yielded a mass of $2.12\,M_\odot$, sufficiently close to the known value. We proceeded to determine the densities of all the clouds in our sample using this method, with values reported in Table~\ref{tab:densities}.

\subsection{Crossmatching with   \textit{Planck} dust emission \label{sec:Planck}}

The   \textit{Planck} mission measured the total emission intensity across the entire sky. For the Galactic plane and for molecular clouds, the emission intensity primarily comes from interstellar dust. This emission is therefore a useful tracer of column density, and was used for the analysis of giant molecular clouds and their magnetic fields in \cite{Planck_2016}. The emission map along with the all-sky temperature profiles was also converted to a corresponding visual extinction map \citep{Planck_2016} which can be used for both mass estimation as well as average extinction. For Bok globules, however, the primary problem lies with   \textit{Planck}'s resolution; its beam size being as large as $5\arcmin$. This results in most of our clouds occupying fewer than 10 pixels in the   \textit{Planck} all-sky map.

We extracted a resampled map of resolution 0.05 degrees (3 arcminutes) using the \texttt{reproject\_healpix} function in Python, and then increased the image resolution via cubic interpolation. Most maps include both the Bok globule itself along with extinction from the Galactic background, which should be subtracted to isolate the foreground cloud. This is done by extracting a $2\fdg{}5\times{}2\fdg{}5$ region around the Bok globule and subtracting the median emission of this image from each pixel. Wherever this region showed presence of additional high-emission background structures that could potentially distort the median, the region was suitably shrunk until the background could be cleanly seen.

Masses and average surrounding extinctions were calculated with the \textit{Gaia} stellar density-based structures defining the cloud outline, similar to the procedure for 2MASS extinction maps. Four of our clouds (BHR 59, 133, 138 and 149) lie very close to the Galactic plane in latitude, and also at a longitude near to 0, causing the Galactic disk dust emission to dominate over the cloud in the foreground. The median is also dominated by the Galactic background as a result, and becomes very high irrespective of the surrounding region size, and median subtraction causes a significant number of pixels to assume negative values of extinction. For these four cases   \textit{Planck} results (both mass and surrounding extinction) are not used any more.

We test the mass determination method again on Barnard 68, and find an excellent agreement of the literature mass of $2.1 M_\odot$ and our   \textit{Planck} derived mass of $2.16 M_\odot$. Additionally, we also find a good agreement to within a factor of 2 in almost all cases between   \textit{Planck} and 2MASS derived values of masses and surrounding extinctions (see Table~\ref{tab:full-properties}).

\begin{figure}\includegraphics[width=\columnwidth]{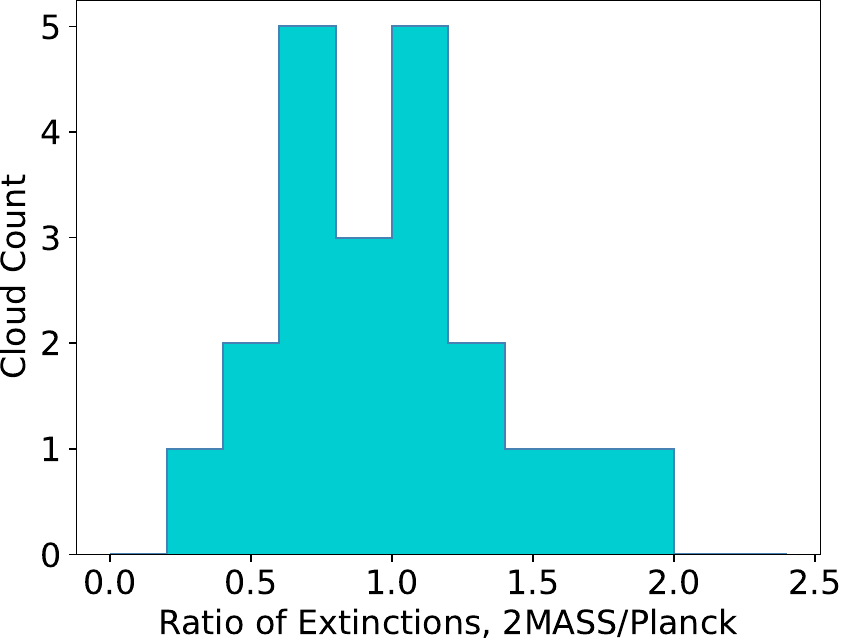}
    \caption{Histogram of the ratios of mean extinction $\overline{A_V}$ obtained from using 2MASS colours to   \textit{Planck} emission maps. The distribution is centred at $\sim 1$, with only one cloud lying outside the region where the ratio is $<0.5$ or $>2$ (corresponding to a difference > factor of 2) }
    \label{fig:av_ratios}
\end{figure}

The extinction estimated using   \textit{Planck} maps are shown in Table~\ref{tab:full-properties}. The distribution of ratios of 2MASS and   \textit{Planck} derived extinctions is given in Figure~\ref{fig:av_ratios}. As can be seen the ratios concentrate around 1 (with a mean of 0.96 and standard deviation of 0.39), implying good agreement between the values derived by two completely different methods (extinction vs emission) in two different wavelength regimes and different physical processes.

We choose to use the 2MASS values of masses and extinctions in our analysis for uniformity, as the high background emission in the case of few clouds in the   \textit{Planck} maps renders the median subtracted extinction maps from   \textit{Planck} unusable for those cases. 2MASS does not have a similar problem as its lower limiting magnitude and near-infrared wavelength regimes means that it only probes the much nearer Bok globules themselves, and not the larger background structures farther away that appear in   \textit{Planck}.

\subsection{Polarization Efficiency}
\label{PE-ext}
The efficiency with which grains align deep into clouds has been studied by analyzing  the fractional polarization per unit visual extinction (known as polarization efficiency) as a function of extinction \citep{Goodman_1992}.  

We use \textit{Gaia} derived $A_G$ values, which can be converted to $A_V$ by multiplying it by a factor of $1/0.789 \approx 1.267$ as given by \cite{Wang_2019}. All stars in the polarization dataset having SNR below 3 i.e. $P/\delta P \leq 3$ were removed, where $P$ and $\delta P$ are the polarization percentage and its corresponding uncertainty as measured at the Observatório do Pico dos Dias, respectively. Crossmatching with \textit{Gaia} stars was done within a 1\arcsec radius of each star with polarization data. Since we only searched for \textit{Gaia} stars with a known $A_G$ value, there were no cases where there were $>1$ crossmatches in the 1\,\arcsec\, radius. Any star closer than the distance of the Bok globule itself i.e. part of foreground were removed, although in most cases there were no such stars. 

Unlike the global extinction maps in the region around the globule, we do not use 2MASS colours for this analysis. The difference in both analyses is justified by two main reasons.

\begin{itemize}
    \item For polarization efficiency calculations, we need the extinction values of individual stars. Using 2MASS colours in such cases increases the uncertainty in $A_V$ because the $A_V = 15.9 \times (H-K-0.13)$ assumes the star's intrinsic colour to be $0.13$, which can vary between $0$ and $0.25$, and lead to up to 2 magnitudes of error in $A_V$. \textit{Gaia} $A_G$ values are calculated using the star's physical properties, parallax and observed colours in its three bands and has a much lower error. 

    \item For extinction maps, the high error in individual stellar extinctions from 2MASS is not as large a problem since the colour is averaged over several stars in a pixel. Further, any analysis (like DCF and mass/column density estimation using average extinction) uses the bulk average of several or all of those pixels, further reducing the overall error. 2MASS, however, offers a key advantage over \textit{Gaia} of being able to probe the higher extinction regions too, being in near-IR. In fact, the highest extinction central regions in 2MASS correspond to empty regions in \textit{Gaia}, rendering its $A_G$ values unusable for determining  extinctions in the denser regions of the core. 
\end{itemize}

After the $A_V$ is determined using the \textit{Gaia} $A_G$, we find the polarization efficiency as $P/A_V$ where $P$ denotes the polarization percentage. We remove stars with $A_V < 0.2$ as they have high relative errors, which propagate into high errors in polarization efficiency. 

We then pick a particular cloud, BHR 121, which has a relatively large number of data points with \textit{Gaia $A_G$} crossmatches, and attempt to see if dust grain alignment modelling can explain the observed values (see Section~\ref{sec:meth_grains}). Since this modeling effort requires us to acquire meaningful constraints on the cloud structure, temperature structure and radiation field and thus customize these models to individual sources, we refrain from conducting a complete analysis on all sources. A dedicated effort on this topic would follow.

\subsection{Quantifying Magnetic Field Strength }\label{sec:meth_B}
We use the Davis-Chandrasekhar-Fermi (DCF) method, as detailed in \cite{Chandrasekhar_Fermi_1953} to find the magnetic field strength from the projected plane-of-sky polarization vectors as follows:
\begin{subequations}
\begin{align}
B_{\rm{pos}} (\mu G) &= f \sqrt{4\pi \left(\frac{\rho}{1\ \mathrm{g\ cm}^{-3}}\right)} \frac{\delta v}{1 \ \mathrm{cm\ s}^{-1}} \left( \frac{\delta \phi}{1\ \mathrm{rad}} \right)^{-1} \\
B_{\rm{tot}} &= \frac{4}{\pi}  B_{\rm{pos}}
\end{align}
\end{subequations}
Here $\rho$ is the mass density of the region we are probing the clouds in g cm$^{-3}$, $\delta v$ is the velocity dispersion in cm s$^{-1}$, and $\delta\phi$ is the polarization angle dispersion in radians. The factor $f=0.5$ is taken from studies with synthetic polarization maps \citep{Heitsch_2001, Ostriker_2001}.

We obtained the $\delta v$ values from the observation method as described in \cite{Otrupcek_2000}, using line widths from the $^{13}$CO $(1 \rightarrow{}0)$ transition line, the values given by Tyler Bourke (via private communication). The spectral data used for the calculation of $\delta v$ is obtained from a 43-arcsecond beam centered at the Bok globule.  $\delta\phi$ values are obtained from fitting a Gaussian to the distribution of polarization position angles and taking its standard deviation.

We extrapolate the velocity dispersion using Larson's line-width size relation \citep{Larson_1981}, $\delta v \propto L^{0.38}$ where $L$ is the size of the region (10 \arcmin\ converted to a length).  

The densities $\rho$ can in principle be obtained directly by using the masses obtained from 2MASS and assuming an ellipsoidal structure and using the structural parameters as obtained from \textit{Gaia}. However, this approach delivers a density estimate for the dense inner regions of the Bok globules, whereas the polarization data probes the outer, less dense regions. We therefore use a length scale equivalent to $10\arcmin$ in our density estimates, as described below.

The values of $n_\mathrm{H_2}$ i.e. the number density obtained by using masses and structures directly were on the order of $n_\mathrm{H_2} = 10^4-10^6~\rm{}\,cm^{-3}$, clearly very high values where optical polarization data cannot be obtained due to high extinction. To infer the actual density $n_\mathrm{H_2}$ of the less dense outer regions where the polarization data comes from, we used the 2MASS maps weighed with the \textit{Gaia} star densities in the $10' \times 10'$ square centered on the Bok globule. To convert the column densities and masses to volume densities, we assume the third axis (along line-of-sight) to also be equal to the linear length equivalent of $10'$ projected at a distance of the respective cloud from Earth, taken from \cite{Racca_2009}. 
This implies that the dense central regions where the star density in optical catalog is 0, is excluded from the density calculations and the outer regions are probed. These returned densities $n_\mathrm{H_2} = 10^2-10^4$ cm$^{-3}$, in agreement of what is expected of the regions (see for example \citealt{Kandori_2020_B68}, \citealt{Kandori_2020_BHR71}). The corresponding column density values, $N_\mathrm{H_2}$ range between $10^{20}$ and $10^{22}$ cm$^{-2}$.
The length scale of 10 arcminutes provides a reasonable lower limit for the densities. To ensure that our globules are not significantly larger than the mapped area in our polarization maps, we examined $1\degree \times 1\degree$   \textit{Planck}-based dust column density maps (Section~\ref{sec:Planck}). These visual inspections reveal that Bok globules are compact features largely unassociated with any extensive large-scale structure. The absence of discernible structures on larger scales suggests it is unlikely that our globules extend significantly beyond the 10 arcminutes along the line of sight. Even so, an extent twice as large along the line of sight would proportionally increase the average density by a factor of 1.2--2.
We then have values of $\rho, \delta v$ and $\delta \phi$ all coming from the same region around the Bok globule, which allows us to use the DCF method to find the plane-of-sky magnetic fields and the total magnetic field using the relations presented above.

We then further use the values and the same equations as outlined in \cite{Pillai_2015} to find the relative strengths of the magnetic field w.r.t.\ the turbulent motion of the gas and the self-gravity of the cloud using the following two metrics.

\begin{itemize}
    \item The Alfv\'en Mach number $\mathcal{M}_A$ that finds the relative strength of magnetic field and turbulence, given by 
    \begin{equation}
    \mathcal{M}_A = \sqrt{3} \sigma_v / v_A \label{eq:machnumber} \end{equation}
    where $v_A$ is the Alfv\'en velocity given by $B_{\rm{}tot}/\sqrt{4\pi \rho}$. Substituting gives $\mathcal{M}_A \propto \delta \phi$. $\mathcal{M}_A$ less than 1 signifies that the force due to magnetic field dominates over that of turbulent force, and is therefore in sub-Alfv\'enic regime. Similarly, $\mathcal{M}_A>1$ is the super-Alfv\'enic regime.

    \item The mass to magnetic flux ratio can be compared to a critical value. The formula is given by 
    \begin{equation}
     \frac{(M/\Phi_B)}{(M/\Phi_B)_{\mathrm{cr}}} = 0.76 \left( \frac{\braket{N_\mathrm{H_2}}}{10^{23} \mathrm{cm}^{-2}} \right) \left( \frac{B_{\rm{}tot}}{1000 \mu G} \right)^{-1} \label{eq:mbratio}
    \end{equation}
    where $\braket{N_\mathrm{H_2}}$ is the average column density of the region we are probing. In this case again the stellar density maps are used to do a weighted average over the less dense regions of the clouds.
    If this ratio is less than 1, then magnetic field dominates over the gravity of the system (sub-critical). A value $>1$ signifies that gravity is dominant, and is thus super-critical.
\end{itemize}

\subsection{Quantifying Error Measures}

\begin{figure}
	\includegraphics[width=\columnwidth]{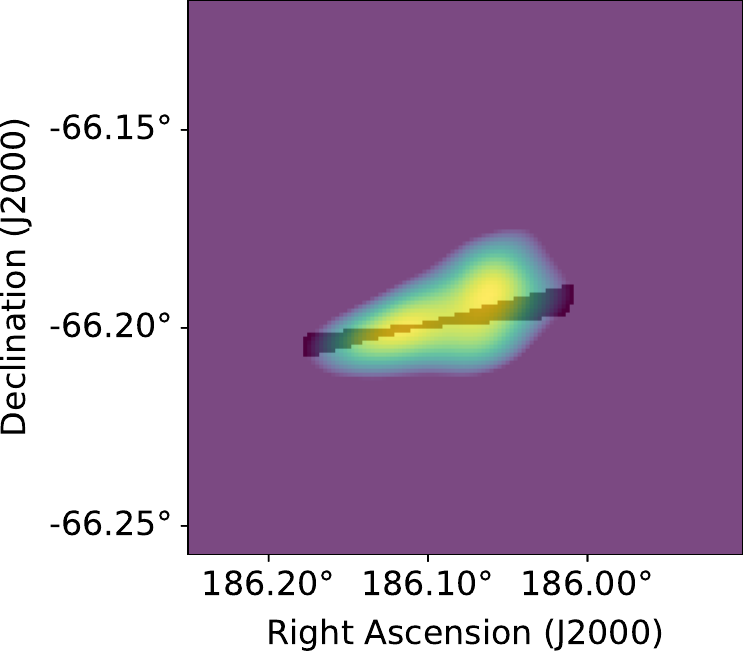}
    \caption{Visual representation of quantifying the error in cloud angle - the red sweep represents the angle from the major axis over which the length of the diameter is $\geq 95\%$ of the major axis length. The angle of the sweep is taken as a measure of the cloud angle error. The image is cropped to the central region where the cloud lies.}
    \label{fig:ellipse-errors}
\end{figure}

Having metrics for errors was essential to arrive at reliable conclusions for the cloud properties. We needed to quantify errors for the following quantities.

\begin{itemize}
    \item \textbf{Distances} - \cite{Racca_2009} does not have error margins on the distance, however their distance bins are at most 25~pc wide in all cases, which we henceforth adopted as the error on the distances of the individual clouds.

    \item \textbf{Velocity Dispersions} - The data obtained from Tyler Bourke did not have error measures for the velocity dispersions; however their channel spacing for the CO velocity dispersions was $0.1~\rm{}km\,s^{-1}$ along the line-of-sight. We henceforth adopt an error upper limit of $0.2~\rm{}km\,s^{-1}$ on the velocity dispersions of individual clouds.

    \item \textbf{Mean and Sigma of Polarization Angle (PA)} - By normal statistics, the error margin on each of these was $\delta\phi / \sqrt{N}$, where $\delta\phi$ is the Gaussian-fit standard deviation on the star sample and $N$ is the sample size.

    \item \textbf{Angle of Cloud Filament} - Since this was obtained purely from images themselves generated via star counts and heuristic convolution and binning, an unambiguous quantification is not possible. This quantity is important as it is also a measure of how close to being highly elliptical (filamentary) a cloud is, or how close to being nearly spherical. 
    
    For each cloud with the ellipse identified, we draw a line through the centre on both sides till it touches the boundary of the elliptical structure. The angle from the major axis of the ellipse at which this line acquires a length equal to $95\%$ of length of major axis, is taken to be the error in the dominant angle of the ellipse. The value of $95\%$ was also chosen by inspection; any lower values gave absurdly large values for the angle error and any higher values did not distinguish well between highly elongated vs spherical clouds due to small non-uniformities on the structure boundary (no cloud boundary was shaped like a smooth ellipse) which was used to find the angle error. 

    This angle error is also the dominant source of error for the relative angle between the cloud and mean polarization which is detailed in the results section.

    \item \textbf{Mass of Cloud (M)} - While not used directly anywhere in further calculations, determination of mass using the structure identification and extinction determination methods detailed above was an important outcome of this work. Hence as an additional check, we found the masses from both 2MASS maps as well as   \textit{Planck} emission (converted to extinction) maps. We find a good agreement of both the measures to within a factor of 2 in all cases except the four outlined earlier, and proceed by assuming the average of the two values to be the true mass of the cloud. Further, we adopt a standard value of error as a factor of 2, which in most cases overestimates the actual error but does not significantly alter the analysis.

    \item \textbf{Average Extinction in Low-Density Surroundings (Weighted $A_V$)} - This is again derived from images binned from 2MASS photometric data but devoid of any heuristic cuts (like $A_V > 3$). The weighted average extinction can incur errors from mainly two sources: stellar densities of \textit{Gaia} used for weighing, and the extinction derived from stellar passband magnitudes from 2MASS, which have average measurement errors from 0.15-0.2 magnitudes, which with the formula given earlier in \cite{Racca_2009} translates to an error in $A_V$ of about 4. We however note that the RA-Dec measurements of stars in \textit{Gaia} astrometry are extremely accurate compared to this and hence stellar density errors can effectively be neglected. However, the stellar densities by themselves cannot be claimed to act as good weights for the extinction maps, hence we do not have a strictly quantifiable error measure on the average extinction from 2MASS. Hence the same extinctions are then also derived with the   \textit{Planck} maps, but that individually suffers from the strong galactic disk background contamination talked about in the previous section which is estimated by subtracting the median extinction from the entire map. The same stellar density based weighing and averaging is used for these maps too, and we find a good agreement between the 2MASS and   \textit{Planck} measures within a factor of 2, excepting those four clouds with strong galactic background. We adopt a factor of 2 error in the average extinction as well, and the actual average extinction to be the mean of the two values obtained from   \textit{Planck} and 2MASS where both are available, and only 2MASS else. Our results remain robust even to this relatively large error on weighted $A_V$.
\end{itemize}
All other quantities are computed using the above main quantities, so errors for the derived quantities was computed using standard Gaussian error propagation methods.

\subsection{Magnetic Field Position Angle Distribution}
\label{sec:pa-dist}

 In Figures~\ref{fig:pa-distribution-1}--~\ref{fig:pa-distribution-2}, we present the PA (polarization angle) distribution for all the clouds in our sample. In the DCF method, the dispersion of the polarization position angles is the key parameter used to estimate the magnetic field strength.  A strongly ordered magnetic field distribution would appear as a Gaussian distribution with a well-defined mean and small angle dispersion, while a weak or randomly oriented field would have a significant dispersion.
Several numerical MHD simulations have investigated the uncertainty of the DCF approach in determining field strength \citep{Ostriker_2001, Heitsch_2001} and derived correction factors based on the angle dispersion. The correction factor of 0.5 used in this work \citep{Ostriker_2001} is valid only for B-field angle dispersions, ${\delta \phi < 25^\circ}$. In our sample of 21 Bok globules, we find 15 to be consistent with having a well-ordered magnetic field, as shown in Figures~\ref{fig:pa-distribution-1}--~\ref{fig:pa-distribution-2}. Two of the clouds, BHR 44 and BHR 145, appear to have randomly distributed fields. Upon visual inspection, the polarization vectors in these clouds appear to trace shell-like structures around the clouds, and attempts to fit a Gaussian result in a poor fit, with dispersions $\geq 50^\circ$, leading to their exclusion from the analysis.

Additionally, four more globules (BHR 53, 59, 148, and 149) show two distinct peaks in the PA distribution histogram. We fit a double Gaussian to account for the bimodal distribution. The two distinct modes are likely influenced by intervening dust from two separate clouds. To confirm this, determining the distances to the stars belonging to each of the two modes is necessary.

\begin{table*}\centering
\begin{tabular}{|c|c|c|c|c|} \hline
\thead{Name} & \thead{First Component \\ PA (Mean $\pm$ Dispersion)} & \thead{Second Component \\ PA (Mean $\pm$ Dispersion)} & \thead{First Component Distance\\ (Mean $\pm$ Dispersion) in pc} & \thead{Second Component Distance \\ (Mean $\pm$ Dispersion) in pc} \\ \hline
BHR 53 & 190.77 $\pm$ 11.8 & 151.0 $\pm$ 16.2 & 2096 $\pm 1254$ & 3218 $\pm$ 1688 \\
BHR 59 & 136.05 $\pm$ 18.1 & 64.52 $\pm$ 27.1 & 2074 $\pm$ 1328 & 3493 $\pm$ 1125 \\
BHR 148 & 34.04 $\pm$ 8.1 & 175.5 $\pm$ 36.9 & 1448 $\pm$ 641 & 2689 $\pm$ 1194 \\
BHR 149 & 33.42 $\pm$ 12.9 & 128.86 $\pm$ 16.5 & 1395 $\pm$ 534 & 3176 $\pm$ 907 \\

\hline
\end{tabular}
\caption{Properties of the clouds having bimodal distribution. First component refers to the nearer one, which is used for latter analysis.}\label{tab:bimodal}
\end{table*}

\begin{figure}
	\includegraphics[width=\columnwidth]{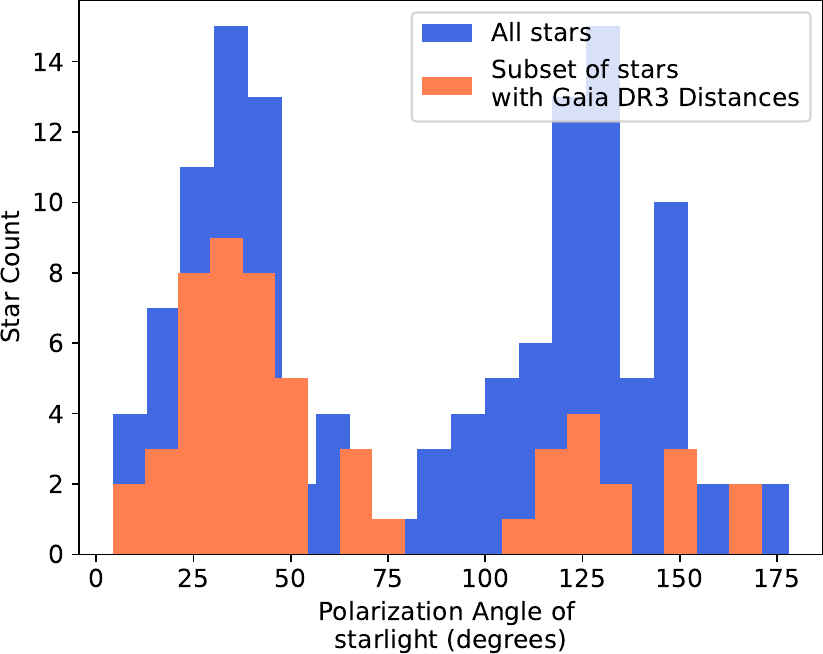}\\
 \includegraphics[width=\columnwidth]{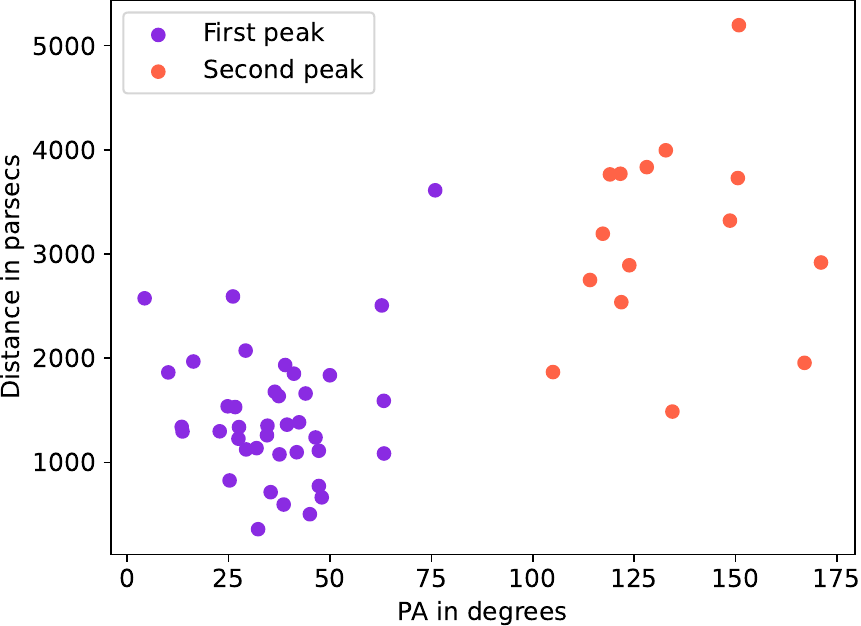}
    \caption{Bimodal distribution of angles in case of BHR 149. Stars belonging to the two different angular components also form two distinct clusters in distance. The parameters corresponding to the nearer set of stars in distance are used for further analysis}
    \label{fig:bimodal}
\end{figure}

We use our \textit{Gaia} crossmatched stars for sources belonging to each of the two modes to see if they show a clear distinction in distance. For three of the four cases, we find the average distances of the two groups (with two different polarization angles) to have a statistically significant ($> 1\sigma$) difference, as shown for an example in Fig~\ref{fig:bimodal}. As any stars whose \textit{Gaia}-derived distances placed them in front of the target cloud were removed already, both the distance modes are still behind the Bok globule in all cases. The means and standard deviations of each polarization component and the corresponding distances given in Table~\ref{tab:bimodal}. Since all the Bok globules in our sample lie at distances of 200 to 700 pc, we conclude that the polarization angle of the nearer set of stars behind the cloud is representative of the true polarization in the vicinity of the Bok globule. The farther component is likely affected by other intervening clouds in the line-of-sight, unrelated to our target globule and hence discarded for further analysis. One of the four clouds, BHR 53, did not have a clear $1\sigma$ difference in the distance of the stars belonging to the two groups of polarization angles, but we still take the one with the nearer distance for further analysis for uniformity.

Additionally, we notice that in all of the four clouds, the stars belonging to the two different PA components have mean polarization fractions that are similar at the $1\sigma$ level.

With the final components taken from these, and after discarding BHR 44 and 145, we have a clean set of polarization position angle distributions with all of them having a dispersion less than $25^\circ$.

\section{Results}\label{sec:results}

\begin{table}\centering
\begin{tabular}{|c|c|c|c|c|} \hline
Name & RA (J2000) & Dec (J2000) & $\delta v$ (km s$^{-1}$) & Star-forming \\ \hline
BHR 16  &  08 05 26 &  -39 08 54 & 0.8 & No \\
BHR 34  &  08 26 34 &  -50 39 54 & 0.7 & Yes \\
BHR 44  &  09 26 19 &  -45 11 0 & 0.8 & No \\
BHR 53  &  09 28 47 &  -51 36 42 & 1.2 & No \\
BHR 58  &  10 49 00 &  -62 23 06 & 1.7 & Yes \\
BHR 59  &  11 07 07 &  -62 05 48 & 0.9 & No \\
BHR 74  &  12 22 09 &  -66 27 06 & 0.4 & No \\
BHR 75  &  12 24 13 &  -66 10 42 & 0.4 & No \\
BHR 111  &  15 42 20 &  -52 49 06 & 1.1 & No \\
BHR 113  &  16 12 43 &  -52 15 36 & 0.9 & No \\
BHR 117  &  16 06 18 &  -45 55 18 & 0.6 & Yes \\
BHR 121  &  16 58 42 &  -50 35 48 & 1.4 & Yes \\
BHR 126  &  16 04 29 &  -39 37 48 & 0.4 & Yes \\
BHR 133  &  16 46 45 &  -44 30 48 & 0.4 & No \\
BHR 138  &  17 19 36 &  -43 27 06 & 1.0 & Yes \\
BHR 139  &  17 20 45 &  -43 20 30 & 1.2 & Yes \\
BHR 140  &  17 22 55 &  -43 22 36 & 1.7 & Yes \\
BHR 144  &  16 37 28 &  -35 13 54 & 0.9 & No \\
BHR 145  &  17 48 01 &  -43 43 12 & 0.8 & No \\
BHR 148  &  17 04 26 &  -36 18 48 & 1.1 & Yes \\
BHR 149  &  17 04 27 &  -36 08 24 & 1.0 & Yes \\

\hline
\end{tabular}
\caption{Properties of Bok Globules obtained from archival data. The velocity dispersion $\delta v$ noted here is for a beam width of 43\arcsec. It is scaled for 10 arcminutes for the analysis, using Larson's relation}\label{tab: }
\end{table}

\begin{table*}
\begin{adjustbox}{width=2.4\columnwidth,center}
\begin{tabular}{|c|c|c|c|c|c|c|c|c|c|c|}  \hline
Name & \thead{Relative \\ Angle (deg)} & \thead{PA Dispersion \\ (deg)} & \thead{2MASS \\ Mass (M$_\odot$)} & \thead{  \textit{Planck} \\ Mass (M$_\odot$)} & \thead{Weighted $A_V$ \\ (2MASS)} & \thead{Weighted $A_V$ \\ (  \textit{Planck})} & \thead{Density $n_\mathrm{H_2}$ \\ $10^3$  cm$^{-3}$} & \thead{Plane-of-Sky \\ Magnetic Field \\ ($\mu$G)} & \thead{Mach Number \\ $\mathcal{M}_A$} & \thead{$\frac{(M/\Phi_B)}{(M/\Phi_B)_{\mathrm{cr}}}$} \\ \hline

BHR 16  &  28.9 $\pm$ 9.8 & 19.6 & 23.4 & 43.9 & 2.1 & 3.3 & 0.7 $\pm$ 0.4 & 37 $\pm$ 17 & 0.92 $\pm$ 0.07 & 0.4 $\pm$ 0.39 \\
BHR 34  &  38.2 $\pm$ 20.1 & 18.4 & 27.6 & 23.4 & 1.4 & 0.8 & 0.4 $\pm$ 0.1 & 24 $\pm$ 11 & 0.86 $\pm$ 0.05 & 0.41 $\pm$ 0.39 \\
BHR 44  &  12.1 $\pm$ 8.4 & 49.9 & 26.8 & 43.4 & 2.4 & 2.2 & 0.8 $\pm$ 0.3 & \ldots  & \ldots  & \ldots  \\
BHR 53  &  12.3 $\pm$ 16.0 & 11.8 & 53.9 & 64.8 & 1.7 & 1.6 & 0.4 $\pm$ 0.1 & 64 $\pm$ 25 & 0.55 $\pm$ 0.04 & 0.19 $\pm$ 0.17 \\
BHR 58  &  55.6 $\pm$ 9.8 & 12.5 & 4.0 & 3.8 & 0.5 & 0.6 & 0.2 $\pm$ 0.1 & 66 $\pm$ 26 & 0.59 $\pm$ 0.04 & 0.05 $\pm$ 0.05 \\
BHR 59  &  78.5 $\pm$ 4.6 & 18.3 & 20.0 & \ldots & 2.5 & \ldots & 1.1 $\pm$ 0.5 & 54 $\pm$ 23 & 0.86 $\pm$ 0.03 & 0.34 $\pm$ 0.32 \\
BHR 74  &  30.4 $\pm$ 18.1 & 10.6 & 2.2 & 0.6 & 0.6 & 0.9 & 0.3 $\pm$ 0.2 & 23 $\pm$ 14 & 0.5 $\pm$ 0.02 & 0.18 $\pm$ 0.19 \\
BHR 75  &  30.2 $\pm$ 6.5 & 12.0 & 2.7 & 2.3 & 0.7 & 0.9 & 0.4 $\pm$ 0.2 & 23 $\pm$ 14 & 0.56 $\pm$ 0.03 & 0.22 $\pm$ 0.24 \\
BHR 111  &  89.7 $\pm$ 3.0 & 10.6 & 33.0 & 35.0 & 4.2 & 3.1 & 1.8 $\pm$ 0.5 & 145 $\pm$ 60 & 0.5 $\pm$ 0.02 & 0.21 $\pm$ 0.19 \\
BHR 113  &  89.6 $\pm$ 2.7 & 5.4 & 27.5 & 18.4 & 5.5 & 13.4 & 2.9 $\pm$ 2.7 & 296 $\pm$ 134 & 0.26 $\pm$ 0.01 & 0.13 $\pm$ 0.13 \\
BHR 117  &  3.0 $\pm$ 11.3 & 12.6 & 10.1 & 23.1 & 1.3 & 1.1 & 0.5 $\pm$ 0.2 & 37 $\pm$ 18 & 0.59 $\pm$ 0.02 & 0.25 $\pm$ 0.25 \\
BHR 121  &  33.9 $\pm$ 2.9 & 8.3 & 1.9 & 3.8 & 0.2 & 0.6 & 0.1 $\pm$ 0.1 & 43 $\pm$ 17 & 0.39 $\pm$ 0.02 & 0.03 $\pm$ 0.02 \\
BHR 126  &  79.2 $\pm$ 14.3 & 11.2 & 6.8 & 10.0 & 1.9 & 2.2 & 1.2 $\pm$ 0.5 & 41 $\pm$ 25 & 0.53 $\pm$ 0.04 & 0.33 $\pm$ 0.37 \\
BHR 133  &  15.4 $\pm$ 4.8 & 13.7 & 393.2 & \ldots & 6.4 & \ldots & 1.0 $\pm$ 0.4 & 30 $\pm$ 16 & 0.64 $\pm$ 0.02 & 1.51 $\pm$ 1.56 \\
BHR 138  &  21.8 $\pm$ 14.9 & 6.2 & 27.9 & \ldots & 1.4 & \ldots & 0.4 $\pm$ 0.2 & 102 $\pm$ 43 & 0.29 $\pm$ 0.02 & 0.1 $\pm$ 0.09 \\
BHR 139  &  3.9 $\pm$ 14.1 & 7.3 & 26.9 & \ldots & 1.3 & \ldots & 0.3 $\pm$ 0.1 & 102 $\pm$ 42 & 0.35 $\pm$ 0.03 & 0.09 $\pm$ 0.09 \\
BHR 140  &  0.1 $\pm$ 8.3 & 9.6 & 22.8 & 52.0 & 1.1 & 1.2 & 0.3 $\pm$ 0.1 & 102 $\pm$ 37 & 0.45 $\pm$ 0.02 & 0.08 $\pm$ 0.07 \\
BHR 144  &  88.3 $\pm$ 13.5 & 17.6 & 4.4 & 5.0 & 1.2 & 2.2 & 0.7 $\pm$ 0.5 & 46 $\pm$ 22 & 0.83 $\pm$ 0.05 & 0.19 $\pm$ 0.18 \\
BHR 145  &  57.6 $\pm$ 7.4 & 106.5 & 27.2 & 29.9 & 1.1 & 1.4 & 0.3 $\pm$ 0.1 & \ldots & \ldots & \ldots \\
BHR 148  &  48.8 $\pm$ 17.6 & 8.1 & 14.6 & 12.2 & 2.9 & 1.6 & 1.5 $\pm$ 0.3 & 177 $\pm$ 78 & 0.38 $\pm$ 0.03 & 0.12 $\pm$ 0.11 \\
BHR 149  &  9.3 $\pm$ 3.9 & 12.5 & 13.2 & \ldots & 2.6 & \ldots & 1.4 $\pm$ 0.6 & 99 $\pm$ 44 & 0.59 $\pm$ 0.04 & 0.19 $\pm$ 0.18 \\

\hline
\end{tabular}
\end{adjustbox}
\caption{Properties of Bok Globules derived in this work. For the bimodal distributions of BHR 59 and BHR 149, only the angular component of the nearer set of stars is retained. The final mass-to-magnetic-flux ratio is calculated using   \textit{Planck} masses and extinctions. Wherever   \textit{Planck} had high galactic background contamination, 2MASS values are used for the same and the corresponding entries of   \textit{Planck} parameters are left blank. BHR 44 and 145 are listed here for completeness, but not used for analysis as they have high PA dispersions, rendering the DCF method not applicable.}\label{tab:full-properties}
\end{table*}

\subsection{Polarization Efficiency and Extinction}

A power law of the standard form $(P/A_V) = k(A_V)^{-p}$ was fitted to the data of polarization efficiency w.r.t. the extinction $A_V$ as obtained in Section~\ref{PE-ext}. For the four clouds which have a bimodal distribution of position angles (BHR 53, 59, 148, 149) and indicated two separate cloud structures at different distances influencing the polarization (discussed in Section~\ref{sec:pa-dist}), we only used the stars belonging to the nearby component for the power law fitting. We also observed that even if we perform the fitting with all stars instead of just the ones in the nearby component, the best-fit index ($p$) changes in value by less than $0.15$ in all cases. 

We took the index $p$ of the star-forming and starless cores separately and found their variance-weighted mean, where the variance is returned by the \texttt{curve\_fit} function itself.

The star-forming clouds had a mean $\mu_p = 0.75$ and standard deviation $\sigma_p = 0.15$. The starless cores have a mean of $\mu_p = 0.73$ and $\sigma_p = 0.30$. 

There is thus no statistically significant difference in the power-law indices for the star-forming cores and the starless cores.

\subsection{Modelling Grain Alignment Efficiency}
\label{sec:meth_grains}

  To place our power law fits the polarization efficiency above in context, we utilize the DustPOL-py model\footnote{Model: \href{https://github.com/lengoctram/DustPOL-py}{https://github.com/lengoctram/DustPOL-py} \\ A web-interface: \href{https://dustpol-py.streamlit.app}{https://dustpol-py.streamlit.app}} for an isolated cloud (without an internal radiation source, see Tram et al. in prep. for details regarding the model basis) to analyze BHR~121—a cloud with extensive data on starlight polarization. The model is fundamentally based on the radiative torque theory for grain alignment (RAT-A; see \citealt{1976Ap&SS..43..291D,DW1996,2007MNRAS.378..910L,2008MNRAS.388..117H,2015ARA&A..53..501A} for further details) and presumes that grains are entirely aligned with the magnetic fields. The most important parameters in this model are the gas volume density and temperature, the radiation field and the dust temperature, and the grain composition and shape. 
 
\begin{figure}
	\includegraphics[width=\columnwidth]{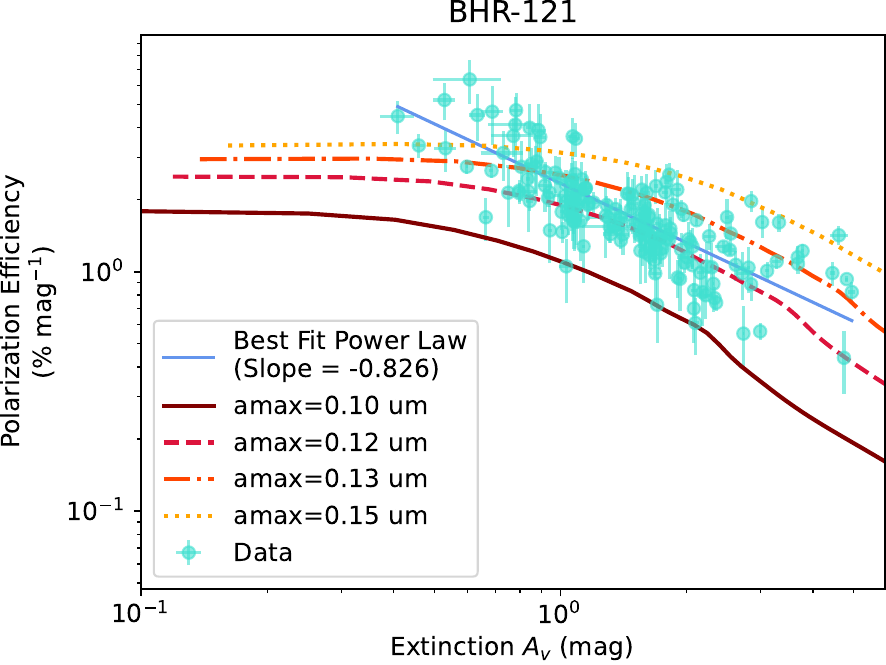}
    \caption{Comparison between the observed data of extinction $A_{\rm v}$ vs polarization efficiency $P/A_{\rm v}$ for BHR 121, with the expected curves obtained by modelling for the same cloud.  Each curve represents a different maximum oblate grain with an axial ratio of 1.4. The standard interstellar radiation field is used. Except for low extinctions, nearly all the data is well-explained by the model.}
    \label{fig:model-data}
\end{figure}

 We utilize a cylindrical configuration incorporating a Plummer density profile across the core of the globule. The gas volume density ($n_{\rm H}=10^{5}\,\rm cm^{-3}$) and changes as a function of the distance ($r$) from the center following $n_{\rm H} \sim r^{-2}$. Our cloud is embedded in the standard ISRF with the mean wavelength of 1.3$\,\mu$m, and the surrounding envelope temperature of 15\,K. The local radiation field, mean wavelength and dust temperature within the cloud are computed through the radiative process.
We make use of the oblate ASTRODUST grain composition \citep{Hensley_2023} with the axial ratio of 1.4. The grain size is varied to investigate the impact of the largest grain size on polarization efficiency across different wavelengths. Figure~\ref{fig:model-data} presents the best models that well cover our data.

As seen in Fig~\ref{fig:model-data}, $93\%$ of our observed data lies on or between the four model curves for different grain sizes. For $A_V \gtrsim 0.7$, the observed polarization efficiencies can be well explained by the typical radiation fields and dust grain sizes found in such environments. At lower extinctions, the polarization efficiencies are systematically higher than the model predictions, implying that they are still aligned with, and hence are a probe of, the local magnetic field, but additional physics like a stronger radiation field or more elongated grains might be needed to explain the higher polarization efficiencies observed. 

\subsection{Relative Angles between Cloud and Field}
To determine the relative orientation of clouds and magnetic fields, we use the Rolling Hough Transform algorithm (section~\ref{sec:rht}) to calculate the dominant orientation of each cloud. For the magnetic field, we calculate the mean angle of polarization vectors by fitting a Gaussian distribution to their position angle  data. The relative angle is then the difference between these two values, computed modulo \(180^\circ\).

\begin{figure}
        \includegraphics[width=8cm]{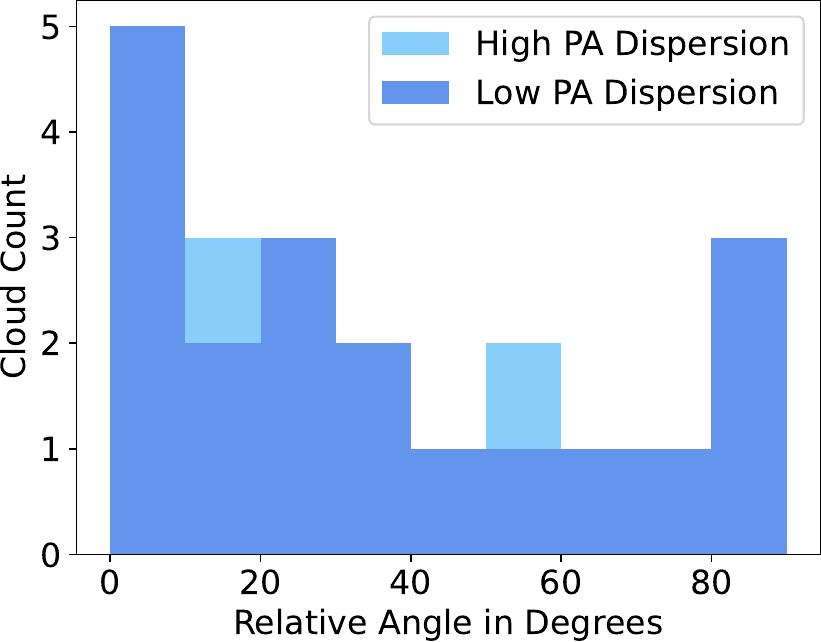} \\
        \includegraphics[width=8cm]{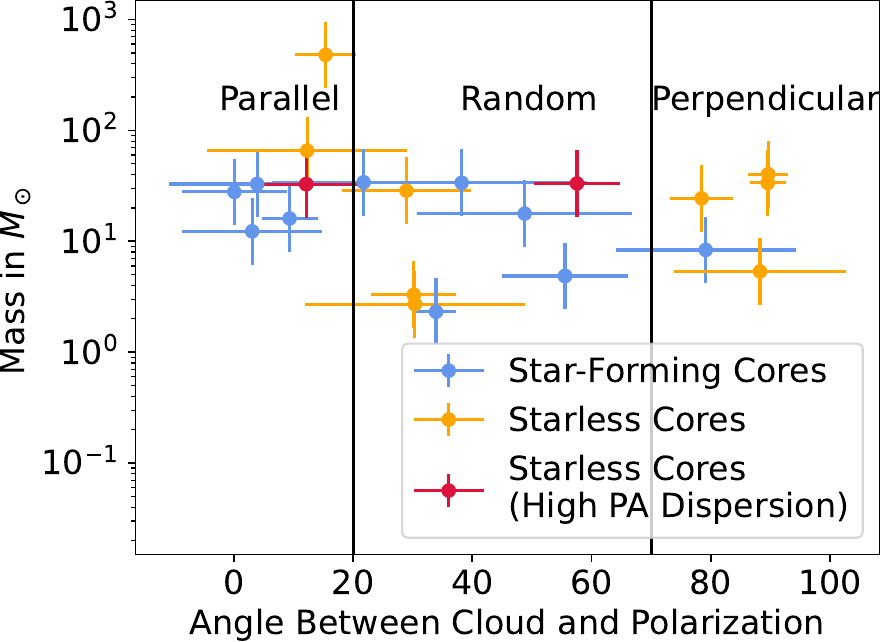}
    \caption{Distribution of relative angles for all clouds, including those with $\delta \phi \geq 25^\circ$, plotted against mass as well as a distribution histogram}
    \label{fig:Orientations}
\end{figure}

Since there is an inherent uncertainty of approximately \(10^\circ\) (corresponding to our 3\,$\sigma$ cutoff on polarization angles) in each angle measurement, we construct a histogram of the relative angles using \(10^\circ\) bins. As shown in the upper panel of Figure~\ref{fig:Orientations}, this allows us to analyze the distribution of relative orientations more robustly.

Additionally, we examine potential correlations between the relative angles and the cloud mass, as derived from 2MASS observations. This step helps us investigate whether the relative orientation depends on cloud mass or follows a specific trend. The results are shown in the lower panel of Figure~\ref{fig:Orientations}.

To ensure the reliability of the analysis, we exclude the following clouds:
- BHR 44 and BHR 145. These clouds exhibit a polarization PA dispersion greater than \(25^\circ\), which means the Gaussian fitting does not yield a clear peak. The resulting mean PA is unreliable and likely random. For BHR 138, the \textit{Gaia} stellar density structure map shows a nearly circular shape, which lacks a distinct filamentary structure necessary for defining a dominant orientation.

After these exclusions, we proceed with the analysis using a final sample of 18 clouds.

\begin{figure}
        \includegraphics[width=8cm]{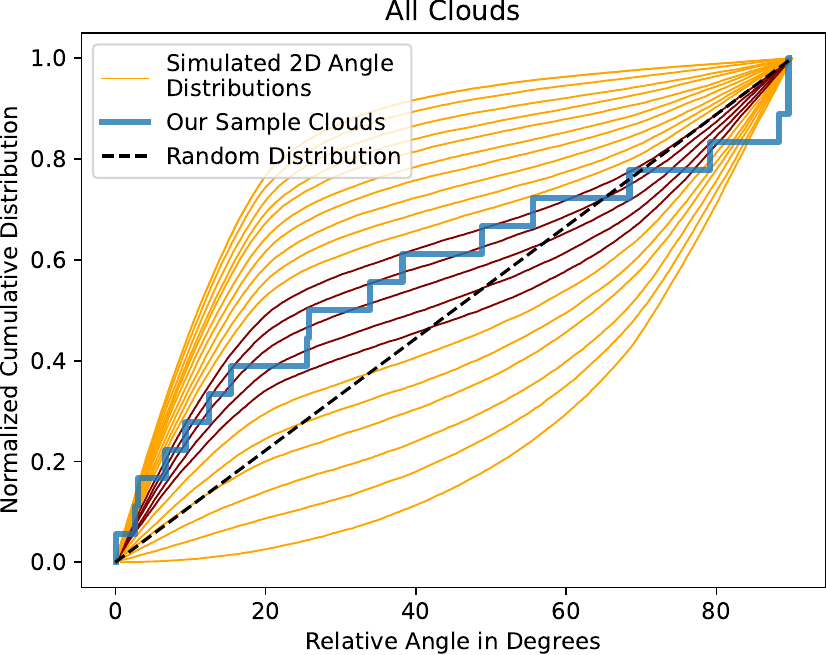} \\
        \includegraphics[width=8cm]{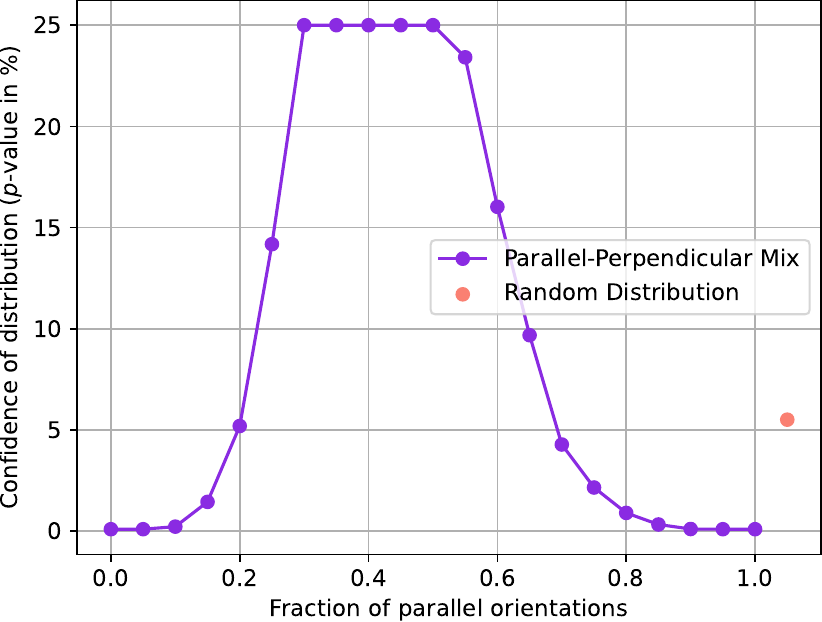}
    \caption{Cumulative distribution of our sample's relative angles, compared against in-sky projected angles of simulated 3D vector pairs. The top panel shows the distributions consistent with our observed sample in dark brown. The bottom panel represents the $p$-value or the confidence level of a given distribution (in terms of $x\%$ parallel, $(100-x)\%$ perpendicular true orientations) and another point for random orientation. The $p$-values are clipped at $0.1\%$ and $25\%$.}
    \label{fig:CDF}
\end{figure}

Following the approach in \cite{Stephens_2017},  we generate the cumulative distribution (CDF) of the observed projected angles between our starlight derived B-field orientations and the filament orientation. The observational results are shown as the blue step curve in Figure~\ref{fig:CDF}. To find the actual distribution of 3D angles between the magnetic field vectors and filament structure vectors, we performed Monte Carlo simulations to generate $5\times 10^4$ 3D vector pairs, representing the orientations of the filament axis and the magnetic field respectively. The angle between each vector pair was taken uniformly from $(0^\circ, 20^\circ)$ with probability $\alpha$ (cases where the filament and magnetic fields are parallel to each other), and uniformly from $(70^\circ, 90^\circ)$ with probability $1-\alpha$ (cases where they are perpendicular to each other). The value of $\alpha$ is varied from 0 to 1 in steps of 0.05. Viewing angles are drawn uniformly from a 3D sphere, and the distribution of the projected plane-of-sky angle between each vector pair is calculated. 
We then apply the Anderson-Darling (AD) test \citep{AD_Test} to determine which distribution, from a set of simulated distributions ranging from purely parallel, a mix of parallel and perpendicular, to purely perpendicular relative angles, most accurately fits our data. We also perform the AD test using simulated vector pairs for a random orientation, where their relative angles in 3D are uniformly drawn from the range $(0^\circ, 90^\circ)$. The $p$-value in the AD test quantifies the probability of observing the given data if it were truly drawn from a hypothesized distribution. In the case of a random alignment of true 3D vectors, with a $p$-value of 0.055, we can reject,  at the  $10\%$ confidence level, the hypothesis that the vectors are randomly distributed.

 Fig~\ref{fig:CDF} shows  the results of the simulated distributions as the set of yellow curves and random distribution as the black dashed line. Our CDF is inconsistent with a random alignment of the true 3D vectors (below a $5.6\%$ confidence level), as well as with a purely parallel or perpendicular alignment (below $0.1\%$ confidence level for both). The family of brown curves in Fig~\ref{fig:CDF}, however, agrees well with our observed distribution where a bimodal distribution of a $30\%-70\%$ to $50\%-50\%$ mix of parallel versus perpendicular 3D vector alignments with a $p$-value greater than $25\%$. (The caps of $0.1\%$ and $25\%$ on the $p$-value are imposed by the \texttt{scipy} function used for the test, and essentially represent thresholds where we can confidently accept or reject the respective hypotheses. The p-value for the range ($30\%-50\%$ of parallel alignments) capped at 0.25, thus indicates a significantly better match of the bimodal distribution with the observed distribution than a purely random distribution.

\subsection{Magnetic Field Properties}
\begin{figure}
	\includegraphics[width=\columnwidth]{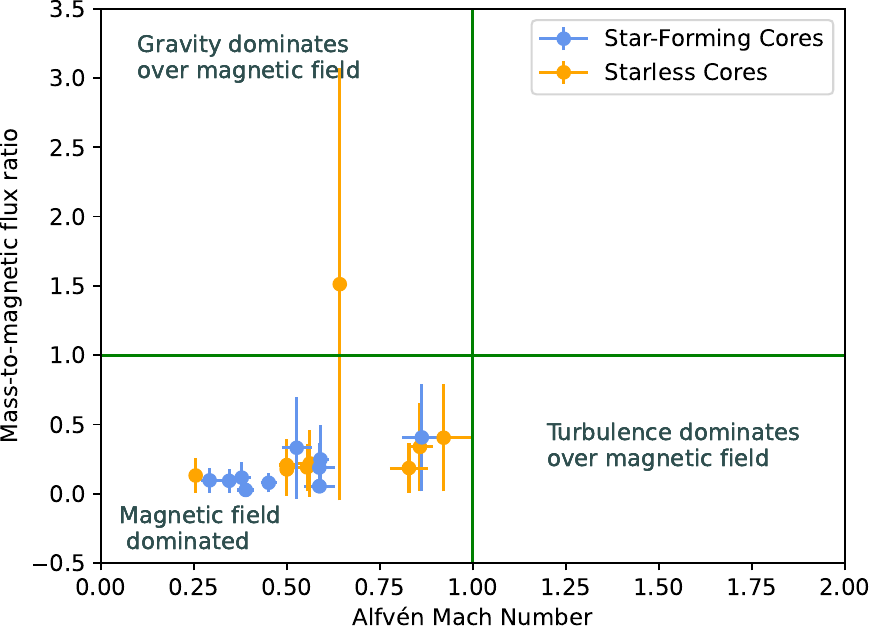}
    \caption{Distribution of mass-to-magnetic flux ratio for all clouds with errorbars. Quantities have been derived using 2MASS masses and extinctions. We note that nearly all our clouds fall into the strongly magnetized regime}
    \label{fig:dcf}
\end{figure}

Here, we present the key findings related to the distribution of magnetic field strengths, mass-to-flux ratios, and Alfv\'{e}n Mach numbers, which provide insights into the magnetic support and turbulence within these Bok Globules, which is also summarized in Fig~\ref{fig:dcf}.

The total magnetic field strength, $( B_{\text{pos}} $), derived using the Davis-Chandrasekhar-Fermi method (refer to Equations~\ref{eq:machnumber}~--\ref{eq:mbratio}), shows a range of values from 23 $\mu$G to 296 $\mu$G.
As shown in Table~\ref{tab:full-properties}, the mass-to-magnetic flux ratios, which quantify the balance between gravitational forces and magnetic support, are found to be in the 0.03--1.51 range  %
All except one globule exhibit mass-to-magnetic  values $<1$. Considering the observational uncertainties,  this sample of Bok Globules appear to be close to being magnetically critical  such that magnetic pressure may provide support on cloud scales against gravitational collapse. That does not necessarily prevent star formation occurring in dense cores within these clouds. The one cloud with a ratio exceeding $1$ is BHR 133, which has an exceptionally high 2MASS-derived mass of $393\, M_\odot$, found to be a starless globule \citep{Racca_2009}. This also happens to be the cloud with the largest distance at 700\,pc, suggesting that distance uncertainties could cause a significant overestimation of the total mass.

The Alfv\'{e}n Mach number, $( \mathcal{M}_A $), derived from the ratio of the turbulent velocity dispersion to the Alfv\'{e}n speed, also exhibits a distribution in the range 0.26 to 0.92. All globules show sub-Alfv\'{e}nic values $( \mathcal{M}_A < 1 $) except the two with high PA dispersion (BHR 44 and 145, where the Alfv\'{e}n Mach number is anyways unreliable as DCF does not work for those clouds). This indicated that magnetic fields are strong enough to influence the cloud dynamics. This suggests that optical polarimetry probes a regime where magnetic forces dominate over those due to turbulence, as can be seen in Fig~\ref{fig:dcf}.

\section{Discussion }\label{sec:disc}

Understanding the role of magnetic fields in the formation and evolution of dense molecular clouds  is essential to advancing our knowledge of star formation  Our optical polarimetric findings towards a sample of 21 Bok Globules  reveal key insights into the magnetic field morphology, and their stability.

\subsection{Grains are efficiently aligned}
A characterization of grain alignment is necessary for accurately analyzing magnetic field strengths and turbulence in the interstellar medium through polarization studies. In one of our Bok Globules with a substantial number of detections (BHR 121), we observe that grain-alignment efficiency declines with increasing extinction, following a power-law index, $<-1$. This finding indicates that, on scales probed by optical polarimetric data, dust grains are mostly aligned with the magnetic field of the cloud. Using a physical model of dust polarization driven by radiative torques toward BHR 121, we confirm that dust grains remain efficiently aligned at a few magnitudes of visual extinction. The polarization efficiency is consistent with previous optical polarimetric study in three Bok Globules \citep{Chakraborty_2014}. Our results also align with previous studies on dense molecular clouds, which suggest that grain alignment can be maintained by radiative torques even in dense environments \citep{Matthews_2000, Henning_2001, Alves_2014, Kandori_2018, Soam_2018, Coude_2019, Pattle_2019, Pillai_2020}.

A comprehensive modeling effort, similar to that conducted for BHR 121, across our full sample may provide deeper insight into the alignment of dust grains, dust grain distribution and the extent to which grain growth or other factors influence alignment efficiency in Bok Globules with varying environmental conditions and density structures.

\subsection{Globules are preferentially extended parallel or perpendicular to the magnetic field} 
An effective way to use polarization measurements to understand the role of magnetic fields is to compare the inferred orientation of the magnetic field with the orientations of elongated molecular cloud structures \citep{Tassis_2009, Li_2013,   Planck_2016}. Bok Globules, as simple and isolated systems, are not significantly affected by large-scale turbulence or nearby star-forming events, making them ideal for probing this relationship. While several studies on magnetic fields in Bok Globules exist \citep{Sen_2000, Ward-Thompson_2009, Bertrang_2014, Choudhury_2022}, large-scale systematic studies on orientation of the Bok Globules versus B-field orientation remain limited. \cite{Ward-Thompson_2009}, found a 40 degree offset between the magnetic field orientation and the short axis in two Bok Globules, while \citet{Chakraborty_2016} find the observed magnetic
field in CB130 to be almost
aligned perpendicular to its minor axis and \citet{Prokopjeva_2014} found magnetic field to be oriented parallel to the Bok Globule's filamentary short axis in CB67. Our survey results for 21 targets is now able to provide context to these case studies. The projected angles between large-scale B-fields and globule long axes do not align purely parallel or perpendicular, showing instead a distribution  consistent with a bimodal alignment, where the alignment is parallel only 30–50\% of the time. Note that we cannot entirely rule out purely random orientation at $\sim 5$\,\% probability. 

\cite{Chen_2014} suggested that gas flows along magnetic field lines, such that its orientation is perpendicular to the main filament. Subsequent theoretical work has further investigated the relative orientation between magnetic fields and column density structures  \citep{Chen_2016, Soler_2017, Seifried_2020} finding a relation for strongly magnetized regimes, with \cite{Seifried_2020} highlighting the impact that projection effects can have in assessing relationships between cloud structures and their magnetic fields, even when they exist.

Observational studies using   \textit{Planck} polarization maps have shown that, in most molecular clouds in nearby Gould Belt regions, B-fields are mostly parallel to filaments at low column density, and a transition occurs from parallel to perpendicular magnetic field alignment with increasing density \citep{Planck_2016}. A similar relation is observed in nearby regions where column density structures are preferentially aligned with magnetic fields \citep{Fissel_2019,Lee_2021,Bij_2024}. This transition from parallel to perpendicular alignment typically occurs at a visual extinction of 3-5\,mag \citep{Planck_2016}.

In our present work, we lack sufficient background stars at or beyond this visual extinction to directly detect such transitions. Deeper near-infrared and longer wavelength polarimetric data that can probe the cores of Bok Globules are needed to confirm the existence of this transition. However, our bimodal distribution results suggest that many of the Bok Globules in our sample exhibit at least moderately strong magnetic fields, indicating they are likely either trans- or sub-Alfvénic.

\begin{figure}
	\includegraphics[width=\columnwidth]{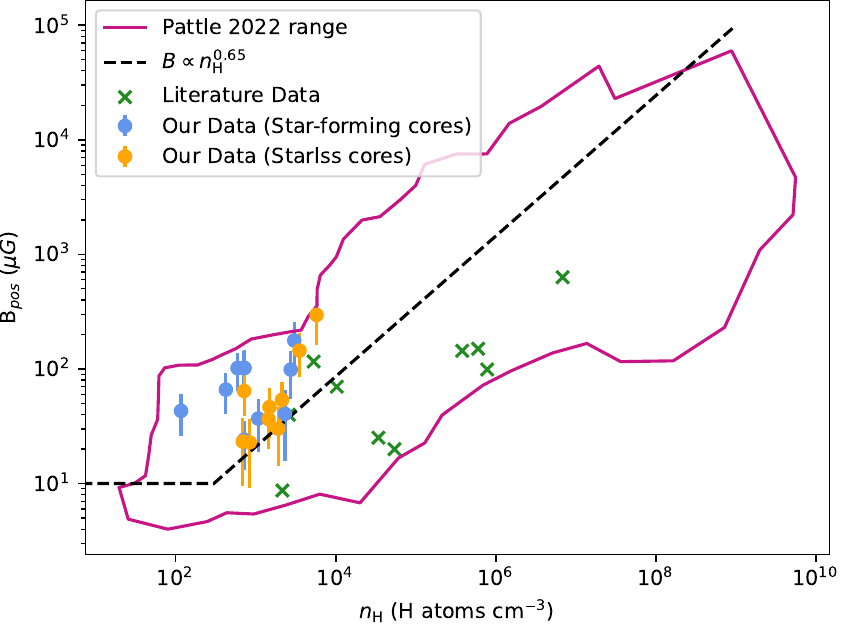}
    \caption{Comparison of B-field and hydrogen density for our clouds vs the DCF based field strengths for different star forming regions by \citet{Pattle_2023} (region marked in red) and specific estimates for Bok Globules (see text). }
    \label{fig:pattle}
\end{figure}

\subsection{Magnetic forces dominate over turbulence and can balance self-gravity}
Our DCF based estimates as shown in Table~\ref{tab:full-properties} further strengthens the case for an important role of magnetic fields. The plane-of-sky magnetic field strength in our sample ranges from 23 to 296\,$\mu$G. Our analysis indicates that the magnetic field within the Bok globule is strongly sub-Alfv\'enic and sub-critical, which has significant implications for its evolutionary path and star formation potential. In the sub-Alfv\'enic regime, magnetic forces outweigh turbulent gas motions, potentially hindering large-scale collapse and stabilizing the globule against gravitational contraction. The sub-critical nature of the globule further suggests that magnetic support prevents gravitational collapse. The free-fall timescale for globule collapse under gravity is approximately 0.1 to 1\,Myr. However, certain globules in our sample are known to host YSOs \citep{Racca_2009}, suggesting that while magnetic support exists on a global scale, it is not sufficient to counteract gravitational forces within the densest cores, enabling localized star formation.

A standard observational test distinguishing strong fields from that of weak fields is the scaling relation between the magnetic field strength and the density it probes: $B \propto n^{\kappa}$ \citep{Crutcher_2012}. In the strong-field model \citep{Mouschovias1999}, the collapse of a molecular cloud driven by ambipolar diffusion primarily compresses gas along magnetic field lines, increasing the density without significantly amplifying the magnetic field strength ($\kappa \le 0.5$). In contrast, weak-field models of cloud collapse, assuming flux freezing, suggest that the magnetic field is too weak to restrict gas flows in any particular direction, leading to a scaling of $\kappa \approx 0.65$ \citep{Mestel1966}. Observational data, based on a compilation of magnetic field strengths measured using Zeeman splitting and their associated densities, align more closely with the weak-field model predictions \citep{Crutcher_2012}. Since our magnetic field strength measurements are based on the DCF method,  in Figure~\ref{fig:pattle}, we compare our findings with an ensemble of DCF-based results from \citet{Pattle_2023}. The reference line represents the theoretically expected magnetic field-density relation ($B \propto n^{0.65}$) for weak fields \citep{Crutcher_2012}. The blue points in Figure~\ref{fig:pattle} show DCF-based Bok Globule observations compiled from the handful of previous studies \citep{Wolf_2003,Vallee_2003, Bertrang_2014, Chakraborty_2016, Das_2016,Choudhury_2019,Kandori_2020_B335, Kandori_2020_B68, Kandori_2020, Myers_2021}. Our results mostly lie above the n$^{0.65}$ relation expected under pure flux-freezing, however it stays within the upper bounds of the literature compilation. They also broadly align with the limited observations of Bok globules available at similar densities. The few existing DCF-based B-field measurements at higher densities (from sub-mm polarimetric data) reveal magnetic fields well below the theoretical reference limit for those densities. The is consistent with gravitational collapse conditions in weak field models \citep{Crutcher_2012}. However a systematic DCF based study of star-forming regions and magnetic field-density relation by \cite{Myers_2021} provide a statistical explanation for the same trend when considering detailed physical properties of these cores under moderately strong magnetic fields.

Our finding makes a key assumption that the DCF method reliably estimates B-field strength. While these approaches are well-established, future research could benefit from multi-wavelength polarimetric observations to more accurately capture B-field morphology across varying cloud depths. Additionally, high-resolution observations—potentially from instruments like ALMA—could offer new insights into small-scale magnetic field variations and further clarify how magnetic fields dissipate to allow star formation that is observed within some of these Bok Globules \citep{Racca_2009}.

\section{Conclusions}\label{sec:conc}
Our study underscores the role of magnetic fields in the formation and stability of Bok globules. Optical polarimetric data from 21 Bok globules probing their envelopes reveal that magnetic fields in these regions are typically strong and sub-Alfv\'enic, meaning magnetic forces largely counteract turbulence, potentially stabilizing the globules against at least large-scale gravitational collapse. Grain alignment modeling for BHR 121 confirms that radiative torques maintain dust grain alignment up to moderate extinction levels, suggesting that optical polarimetry can be a valuable tracer of magnetic field structure in these simple systems. Our findings also suggest that magnetic fields are not purely parallel or perpendicular to globule structures but are in best agreement with a bimodal alignment. The magnetic field strengths, ranging from 23 to 296\,$\mu$G, align with past observations of dense clouds as well as limited studies of Bok Globules. Future studies with deeper infrared polarimetry and higher-resolution observations could better probe these B-field and density interactions, providing more detailed insights into how magnetic support affects star formation within Bok globules.

\section*{Acknowledgements}

TGSP gratefully acknowledges support by the National Science Foundation under grant No. AST-2009842 and AST-2108989 and by NASA award  \#09-0215 
issued by USRA. We thank Dan Clemens and Phil Myers for helpful comments. 
CVR thanks the Brazilian Space Agency (AEB) by the support from PO 20VB.0009 and the Conselho Nacional de Desenvolvimento Científico e Tecnológico - CNPq (Proc:	310930/2021-9).
This paper is based on observations made at the Observatório do Pico dos Dias managed by Laboratório Nacional de Astrof\'\i sica (Brazil).

\section*{Data Availability}
The optical polarization data obtained from the Observatório do Pico dos Dias is available upon reasonable request. All other data (\textit{Gaia, Planck, Herschel} and 2MASS) are obtained from publicly available surveys.



\bibliographystyle{mnras}
\bibliography{bok} 

\begin{thebibliography}{}
\makeatletter
\relax
\def\mn@urlcharsother{\let\do\@makeother \do\$\do\&\do\#\do\^\do\_\do\%\do\~}
\def\mn@doi{\begingroup\mn@urlcharsother \@ifnextchar [ {\mn@doi@} {\mn@doi@[]}}
\def\mn@doi@[#1]#2{\def\@tempa{#1}\ifx\@tempa\@empty \href {http://dx.doi.org/#2} {doi:#2}\else \href {http://dx.doi.org/#2} {#1}\fi \endgroup}
\def\mn@eprint#1#2{\mn@eprint@#1:#2::\@nil}
\def\mn@eprint@arXiv#1{\href {http://arxiv.org/abs/#1} {{\tt arXiv:#1}}}
\def\mn@eprint@dblp#1{\href {http://dblp.uni-trier.de/rec/bibtex/#1.xml} {dblp:#1}}
\def\mn@eprint@#1:#2:#3:#4\@nil{\def\@tempa {#1}\def\@tempb {#2}\def\@tempc {#3}\ifx \@tempc \@empty \let \@tempc \@tempb \let \@tempb \@tempa \fi \ifx \@tempb \@empty \def\@tempb {arXiv}\fi \@ifundefined {mn@eprint@\@tempb}{\@tempb:\@tempc}{\expandafter \expandafter \csname mn@eprint@\@tempb\endcsname \expandafter{\@tempc}}}

\bibitem[\protect\citeauthoryear{{Alves} \& {Yun}}{{Alves} \& {Yun}}{1994}]{Alves_1994}
{Alves} J.,  {Yun} J.,  1994, in {Clemens} D.~P.,  {Barvainis} R.,  eds,  Astronomical Society of the Pacific Conference Series Vol. 65, Clouds, Cores, and Low Mass Stars. p.~230

\bibitem[\protect\citeauthoryear{Alves, Lada  \& Lada}{Alves et~al.}{2001}]{Alves_2001}
Alves J.,  Lada C.,   Lada E.,  2001, \mn@doi [Nature] {10.1038/35051509}, 409, 159

\bibitem[\protect\citeauthoryear{{Alves}, {Frau}, {Girart}, {Franco}, {Santos}  \& {Wiesemeyer}}{{Alves} et~al.}{2014}]{Alves_2014}
{Alves} F.~O.,  {Frau} P.,  {Girart} J.~M.,  {Franco} G.~A.~P.,  {Santos} F.~P.,   {Wiesemeyer} H.,  2014, \mn@doi [\aap] {10.1051/0004-6361/201424678}, \href {https://ui.adsabs.harvard.edu/abs/2014A&A...569L...1A} {569, L1}

\bibitem[\protect\citeauthoryear{Anderson \& Darling}{Anderson \& Darling}{1954}]{AD_Test}
Anderson T.~W.,  Darling D.~A.,  1954, \mn@doi [Journal of the American Statistical Association] {10.1080/01621459.1954.10501232}, 49, 765

\bibitem[\protect\citeauthoryear{{Andersson}}{{Andersson}}{2015}]{Andersson_2015}
{Andersson} B.~G.,  2015, in AGU Fall Meeting Abstracts. pp P34B--03

\bibitem[\protect\citeauthoryear{{Andersson}, {Lazarian}  \& {Vaillancourt}}{{Andersson} et~al.}{2015}]{2015ARA&A..53..501A}
{Andersson} B.~G.,  {Lazarian} A.,   {Vaillancourt} J.~E.,  2015, \mn@doi [ARAA] {10.1146/annurev-astro-082214-122414}, \href {https://ui.adsabs.harvard.edu/abs/2015ARA&A..53..501A} {53, 501}

\bibitem[\protect\citeauthoryear{{Bertrang}, {Wolf}  \& {Das}}{{Bertrang} et~al.}{2014}]{Bertrang_2014}
{Bertrang} G.,  {Wolf} S.,   {Das} H.~S.,  2014, \mn@doi [\aap] {10.1051/0004-6361/201323091}, \href {https://ui.adsabs.harvard.edu/abs/2014A&A...565A..94B} {565, A94}

\bibitem[\protect\citeauthoryear{{Bij} et~al.,}{{Bij} et~al.}{2024}]{Bij_2024}
{Bij} A.,  et~al., 2024, \mn@doi [arXiv e-prints] {10.48550/arXiv.2409.03558}, \href {https://ui.adsabs.harvard.edu/abs/2024arXiv240903558B} {p. arXiv:2409.03558}

\bibitem[\protect\citeauthoryear{{Bohlin}, {Savage}  \& {Drake}}{{Bohlin} et~al.}{1978}]{Bohlin_1978}
{Bohlin} R.~C.,  {Savage} B.~D.,   {Drake} J.~F.,  1978, \mn@doi [\apj] {10.1086/156357}, \href {https://ui.adsabs.harvard.edu/abs/1978ApJ...224..132B} {224, 132}

\bibitem[\protect\citeauthoryear{{Bok}}{{Bok}}{1948}]{Bok_1948}
{Bok} B.~J.,  1948, in , Vol.~7, Harvard Observatory Monographs.
p.~53

\bibitem[\protect\citeauthoryear{{Bok} \& {Reilly}}{{Bok} \& {Reilly}}{1947}]{Bok_1947}
{Bok} B.~J.,  {Reilly} E.~F.,  1947, \mn@doi [\apj] {10.1086/144901}, \href {https://ui.adsabs.harvard.edu/abs/1947ApJ...105..255B} {105, 255}

\bibitem[\protect\citeauthoryear{{Bourke}, {Hyland}  \& {Robinson}}{{Bourke} et~al.}{1995}]{BHR_1995}
{Bourke} T.~L.,  {Hyland} A.~R.,   {Robinson} G.,  1995, \mn@doi [\mnras] {10.1093/mnras/276.4.1052}, \href {https://ui.adsabs.harvard.edu/abs/1995MNRAS.276.1052B} {276, 1052}

\bibitem[\protect\citeauthoryear{{Chakraborty} \& {Das}}{{Chakraborty} \& {Das}}{2016}]{Chakraborty_2016}
{Chakraborty} A.,  {Das} H.~S.,  2016, \mn@doi [\apss] {10.1007/s10509-016-2905-y}, \href {https://ui.adsabs.harvard.edu/abs/2016Ap&SS.361..321C} {361, 321}

\bibitem[\protect\citeauthoryear{{Chakraborty}, {Das}  \& {Paul}}{{Chakraborty} et~al.}{2014}]{Chakraborty_2014}
{Chakraborty} A.,  {Das} H.~S.,   {Paul} D.,  2014, \mn@doi [\mnras] {10.1093/mnras/stu761}, \href {https://ui.adsabs.harvard.edu/abs/2014MNRAS.442..479C} {442, 479}

\bibitem[\protect\citeauthoryear{{Chandrasekhar} \& {Fermi}}{{Chandrasekhar} \& {Fermi}}{1953}]{Chandrasekhar_Fermi_1953}
{Chandrasekhar} S.,  {Fermi} E.,  1953, \mn@doi [\apj] {10.1086/145732}, \href {https://ui.adsabs.harvard.edu/abs/1953ApJ...118..116C} {118, 116}

\bibitem[\protect\citeauthoryear{{Chen} \& {Ostriker}}{{Chen} \& {Ostriker}}{2014}]{Chen_2014}
{Chen} C.-Y.,  {Ostriker} E.~C.,  2014, \mn@doi [\apj] {10.1088/0004-637X/785/1/69}, \href {https://ui.adsabs.harvard.edu/abs/2014ApJ...785...69C} {785, 69}

\bibitem[\protect\citeauthoryear{{Chen}, {King}  \& {Li}}{{Chen} et~al.}{2016}]{Chen_2016}
{Chen} C.-Y.,  {King} P.~K.,   {Li} Z.-Y.,  2016, \mn@doi [\apj] {10.3847/0004-637X/829/2/84}, \href {https://ui.adsabs.harvard.edu/abs/2016ApJ...829...84C} {829, 84}

\bibitem[\protect\citeauthoryear{{Choudhury}, {Barman}, {Das}  \& {Medhi}}{{Choudhury} et~al.}{2019}]{Choudhury_2019}
{Choudhury} G.~B.,  {Barman} A.,  {Das} H.~S.,   {Medhi} B.~J.,  2019, \mn@doi [\mnras] {10.1093/mnras/stz1205}, \href {https://ui.adsabs.harvard.edu/abs/2019MNRAS.487..475C} {487, 475}

\bibitem[\protect\citeauthoryear{{Choudhury}, {Das}, {Medhi}, {Pandey}, {Wolf}, {Dhar}  \& {Mazarbhuiya}}{{Choudhury} et~al.}{2022}]{Choudhury_2022}
{Choudhury} G.~B.,  {Das} H.~S.,  {Medhi} B.~J.,  {Pandey} J.~C.,  {Wolf} S.,  {Dhar} T.~K.,   {Mazarbhuiya} A.~M.,  2022, \mn@doi [Research in Astronomy and Astrophysics] {10.1088/1674-4527/ac6f49}, \href {https://ui.adsabs.harvard.edu/abs/2022RAA....22g5003C} {22, 075003}

\bibitem[\protect\citeauthoryear{Clark, Peek  \& Putman}{Clark et~al.}{2014}]{Clark_2014}
Clark S.~E.,  Peek J. E.~G.,   Putman M.~E.,  2014, \mn@doi [The Astrophysical Journal] {10.1088/0004-637x/789/1/82}, 789, 82

\bibitem[\protect\citeauthoryear{{Clemens} \& {Barvainis}}{{Clemens} \& {Barvainis}}{1988}]{Clemens_1988}
{Clemens} D.~P.,  {Barvainis} R.,  1988, \mn@doi [\apjs] {10.1086/191288}, \href {https://ui.adsabs.harvard.edu/abs/1988ApJS...68..257C} {68, 257}

\bibitem[\protect\citeauthoryear{{Clemens}, {Yun}  \& {Heyer}}{{Clemens} et~al.}{1991}]{Clemens_1991}
{Clemens} D.~P.,  {Yun} J.~L.,   {Heyer} M.~H.,  1991, \mn@doi [\apjs] {10.1086/191552}, \href {https://ui.adsabs.harvard.edu/abs/1991ApJS...75..877C} {75, 877}

\bibitem[\protect\citeauthoryear{{Coud{\'e}} et~al.,}{{Coud{\'e}} et~al.}{2019}]{Coude_2019}
{Coud{\'e}} S.,  et~al., 2019, \mn@doi [\apj] {10.3847/1538-4357/ab1b23}, \href {https://ui.adsabs.harvard.edu/abs/2019ApJ...877...88C} {877, 88}

\bibitem[\protect\citeauthoryear{{Crutcher}}{{Crutcher}}{2012}]{Crutcher_2012}
{Crutcher} R.~M.,  2012, \mn@doi [\araa] {10.1146/annurev-astro-081811-125514}, \href {https://ui.adsabs.harvard.edu/abs/2012ARA&A..50...29C} {50, 29}

\bibitem[\protect\citeauthoryear{{Das}, {Das}, {Medhi}  \& {Wolf}}{{Das} et~al.}{2016}]{Das_2016}
{Das} A.,  {Das} H.~S.,  {Medhi} B.~J.,   {Wolf} S.,  2016, \mn@doi [\apss] {10.1007/s10509-016-2966-y}, \href {https://ui.adsabs.harvard.edu/abs/2016Ap&SS.361..381D} {361, 381}

\bibitem[\protect\citeauthoryear{{Davis}}{{Davis}}{1951}]{Davis_1951}
{Davis} L.,  1951, \mn@doi [Physical Review] {10.1103/PhysRev.81.890.2}, \href {https://ui.adsabs.harvard.edu/abs/1951PhRv...81..890D} {81, 890}

\bibitem[\protect\citeauthoryear{{Davis}, {Chrysostomou}, {Matthews}, {Jenness}  \& {Ray}}{{Davis} et~al.}{2000}]{Davis_2000}
{Davis} C.~J.,  {Chrysostomou} A.,  {Matthews} H.~E.,  {Jenness} T.,   {Ray} T.~P.,  2000, \mn@doi [\apjl] {10.1086/312476}, \href {https://ui.adsabs.harvard.edu/abs/2000ApJ...530L.115D} {530, L115}

\bibitem[\protect\citeauthoryear{{Dolginov} \& {Mitrofanov}}{{Dolginov} \& {Mitrofanov}}{1976}]{1976Ap&SS..43..291D}
{Dolginov} A.~Z.,  {Mitrofanov} I.~G.,  1976, \mn@doi [Ap\&SS] {10.1007/BF00640010}, \href {https://ui.adsabs.harvard.edu/abs/1976Ap&SS..43..291D} {43, 291}

\bibitem[\protect\citeauthoryear{{Draine} \& {Weingartner}}{{Draine} \& {Weingartner}}{1996}]{DW1996}
{Draine} B.~T.,  {Weingartner} J.~C.,  1996, \mn@doi [\apj] {10.1086/177887}, \href {https://ui.adsabs.harvard.edu/abs/1996ApJ...470..551D} {470, 551}

\bibitem[\protect\citeauthoryear{{Draine} \& {Weingartner}}{{Draine} \& {Weingartner}}{1997}]{Draine_1997}
{Draine} B.~T.,  {Weingartner} J.~C.,  1997, \mn@doi [\apj] {10.1086/304008}, \href {https://ui.adsabs.harvard.edu/abs/1997ApJ...480..633D} {480, 633}

\bibitem[\protect\citeauthoryear{{Fissel} et~al.,}{{Fissel} et~al.}{2019}]{Fissel_2019}
{Fissel} L.~M.,  et~al., 2019, \mn@doi [\apj] {10.3847/1538-4357/ab1eb0}, \href {https://ui.adsabs.harvard.edu/abs/2019ApJ...878..110F} {878, 110}

\bibitem[\protect\citeauthoryear{{Goodman}, {Jones}, {Lada}  \& {Myers}}{{Goodman} et~al.}{1992}]{Goodman_1992}
{Goodman} A.~A.,  {Jones} T.~J.,  {Lada} E.~A.,   {Myers} P.~C.,  1992, \mn@doi [\apj] {10.1086/171907}, \href {https://ui.adsabs.harvard.edu/abs/1992ApJ...399..108G} {399, 108}

\bibitem[\protect\citeauthoryear{Griffin et~al.,}{Griffin et~al.}{2010}]{Griffin_2010}
Griffin M.~J.,  et~al., 2010, \mn@doi [Astronomy and Astrophysics] {10.1051/0004-6361/201014519}, 518, L3

\bibitem[\protect\citeauthoryear{{Hall}}{{Hall}}{1949}]{Hall_1949}
{Hall} J.~S.,  1949, \mn@doi [Science] {10.1126/science.109.2825.166}, \href {https://ui.adsabs.harvard.edu/abs/1949Sci...109..166H} {109, 166}

\bibitem[\protect\citeauthoryear{{Heitsch}, {Zweibel}, {Mac Low}, {Li}  \& {Norman}}{{Heitsch} et~al.}{2001}]{Heitsch_2001}
{Heitsch} F.,  {Zweibel} E.~G.,  {Mac Low} M.-M.,  {Li} P.,   {Norman} M.~L.,  2001, \mn@doi [\apj] {10.1086/323489}, \href {https://ui.adsabs.harvard.edu/abs/2001ApJ...561..800H} {561, 800}

\bibitem[\protect\citeauthoryear{{Henning} \& {Launhardt}}{{Henning} \& {Launhardt}}{1998}]{Henning_Launhardt_1998}
{Henning} T.,  {Launhardt} R.,  1998, \aap, \href {https://ui.adsabs.harvard.edu/abs/1998A&A...338..223H} {338, 223}

\bibitem[\protect\citeauthoryear{{Henning}, {Wolf}, {Launhardt}  \& {Waters}}{{Henning} et~al.}{2001}]{Henning_2001}
{Henning} T.,  {Wolf} S.,  {Launhardt} R.,   {Waters} R.,  2001, \mn@doi [\apj] {10.1086/323362}, \href {https://ui.adsabs.harvard.edu/abs/2001ApJ...561..871H} {561, 871}

\bibitem[\protect\citeauthoryear{{Hensley} \& {Draine}}{{Hensley} \& {Draine}}{2023}]{Hensley_2023}
{Hensley} B.~S.,  {Draine} B.~T.,  2023, \mn@doi [\apj] {10.3847/1538-4357/acc4c2}, \href {https://ui.adsabs.harvard.edu/abs/2023ApJ...948...55H} {948, 55}

\bibitem[\protect\citeauthoryear{{Hiltner}}{{Hiltner}}{1949}]{Hiltner_1949}
{Hiltner} W.~A.,  1949, \mn@doi [\nat] {10.1038/163283a0}, \href {https://ui.adsabs.harvard.edu/abs/1949Natur.163..283H} {163, 283}

\bibitem[\protect\citeauthoryear{{Hoang} \& {Lazarian}}{{Hoang} \& {Lazarian}}{2008}]{2008MNRAS.388..117H}
{Hoang} T.,  {Lazarian} A.,  2008, \mn@doi [\mnras] {10.1111/j.1365-2966.2008.13249.x}, \href {https://ui.adsabs.harvard.edu/abs/2008MNRAS.388..117H} {388, 117}

\bibitem[\protect\citeauthoryear{{Huard}, {Sandell}  \& {Weintraub}}{{Huard} et~al.}{1999}]{Huard_1999}
{Huard} T.~L.,  {Sandell} G.,   {Weintraub} D.~A.,  1999, \mn@doi [\apj] {10.1086/308022}, \href {https://ui.adsabs.harvard.edu/abs/1999ApJ...526..833H} {526, 833}

\bibitem[\protect\citeauthoryear{{Jones}, {Hyland}  \& {Bailey}}{{Jones} et~al.}{1984}]{Jones_1984}
{Jones} T.~J.,  {Hyland} A.~R.,   {Bailey} J.,  1984, \mn@doi [\apj] {10.1086/162247}, \href {https://ui.adsabs.harvard.edu/abs/1984ApJ...282..675J} {282, 675}

\bibitem[\protect\citeauthoryear{{Kandori} et~al.,}{{Kandori} et~al.}{2005}]{Kandori_2005}
{Kandori} R.,  et~al., 2005, \mn@doi [\aj] {10.1086/444619}, \href {https://ui.adsabs.harvard.edu/abs/2005AJ....130.2166K} {130, 2166}

\bibitem[\protect\citeauthoryear{{Kandori} et~al.,}{{Kandori} et~al.}{2018}]{Kandori_2018}
{Kandori} R.,  et~al., 2018, \mn@doi [\apj] {10.3847/1538-4357/aadb3f}, \href {https://ui.adsabs.harvard.edu/abs/2018ApJ...865..121K} {865, 121}

\bibitem[\protect\citeauthoryear{{Kandori} et~al.,}{{Kandori} et~al.}{2020a}]{Kandori_2020_B68}
{Kandori} R.,  et~al., 2020a, \mn@doi [\pasj] {10.1093/pasj/psz127}, \href {https://ui.adsabs.harvard.edu/abs/2020PASJ...72....8K} {72, 8}

\bibitem[\protect\citeauthoryear{{Kandori} et~al.,}{{Kandori} et~al.}{2020b}]{Kandori_2020_B335}
{Kandori} R.,  et~al., 2020b, \mn@doi [\apj] {10.3847/1538-4357/ab6f07}, \href {https://ui.adsabs.harvard.edu/abs/2020ApJ...891...55K} {891, 55}

\bibitem[\protect\citeauthoryear{{Kandori} et~al.,}{{Kandori} et~al.}{2020c}]{Kandori_2020}
{Kandori} R.,  et~al., 2020c, \mn@doi [\apj] {10.3847/1538-4357/ab7b68}, \href {https://ui.adsabs.harvard.edu/abs/2020ApJ...892..128K} {892, 128}

\bibitem[\protect\citeauthoryear{{Kandori} et~al.,}{{Kandori} et~al.}{2020d}]{Kandori_2020_BHR71}
{Kandori} R.,  et~al., 2020d, \mn@doi [\apj] {10.3847/1538-4357/ab7b68}, \href {https://ui.adsabs.harvard.edu/abs/2020ApJ...892..128K} {892, 128}

\bibitem[\protect\citeauthoryear{{Kane}, {Clemens}, {Leach}  \& {Barvainis}}{{Kane} et~al.}{1995}]{Kane_1995}
{Kane} B.~D.,  {Clemens} D.~P.,  {Leach} R.~W.,   {Barvainis} R.,  1995, \mn@doi [\apj] {10.1086/175693}, \href {https://ui.adsabs.harvard.edu/abs/1995ApJ...445..269K} {445, 269}

\bibitem[\protect\citeauthoryear{{Kauffmann}, {Bertoldi}, {Bourke}, {Evans}  \& {Lee}}{{Kauffmann} et~al.}{2008}]{Kauffmann_2008}
{Kauffmann} J.,  {Bertoldi} F.,  {Bourke} T.~L.,  {Evans} N.~J. I.,   {Lee} C.~W.,  2008, \mn@doi [\aap] {10.1051/0004-6361:200809481}, \href {https://ui.adsabs.harvard.edu/abs/2008A&A...487..993K} {487, 993}

\bibitem[\protect\citeauthoryear{{Klebe} \& {Jones}}{{Klebe} \& {Jones}}{1990}]{Klebe_1990}
{Klebe} D.,  {Jones} T.~J.,  1990, \mn@doi [\aj] {10.1086/115357}, \href {https://ui.adsabs.harvard.edu/abs/1990AJ.....99..638K} {99, 638}

\bibitem[\protect\citeauthoryear{{Koch} \& {Rosolowsky}}{{Koch} \& {Rosolowsky}}{2015}]{Koch_2015}
{Koch} E.~W.,  {Rosolowsky} E.~W.,  2015, \mn@doi [\mnras] {10.1093/mnras/stv1521}, \href {https://ui.adsabs.harvard.edu/abs/2015MNRAS.452.3435K} {452, 3435}

\bibitem[\protect\citeauthoryear{{Lada}, {Lada}, {Clemens}  \& {Bally}}{{Lada} et~al.}{1994}]{Lada1994}
{Lada} C.~J.,  {Lada} E.~A.,  {Clemens} D.~P.,   {Bally} J.,  1994, \mn@doi [\apj] {10.1086/174354}, \href {https://ui.adsabs.harvard.edu/abs/1994ApJ...429..694L} {429, 694}

\bibitem[\protect\citeauthoryear{{Larson}}{{Larson}}{1981}]{Larson_1981}
{Larson} R.~B.,  1981, \mn@doi [\mnras] {10.1093/mnras/194.4.809}, \href {https://ui.adsabs.harvard.edu/abs/1981MNRAS.194..809L} {194, 809}

\bibitem[\protect\citeauthoryear{{Launhardt} \& {Henning}}{{Launhardt} \& {Henning}}{1997}]{Launhardt_Henning_1997}
{Launhardt} R.,  {Henning} T.,  1997, \aap, \href {https://ui.adsabs.harvard.edu/abs/1997A&A...326..329L} {326, 329}

\bibitem[\protect\citeauthoryear{{Launhardt} et~al.,}{{Launhardt} et~al.}{2010}]{Launhardt_2010}
{Launhardt} R.,  et~al., 2010, \mn@doi [\apjs] {10.1088/0067-0049/188/1/139}, \href {https://ui.adsabs.harvard.edu/abs/2010ApJS..188..139L} {188, 139}

\bibitem[\protect\citeauthoryear{{Lazarian} \& {Hoang}}{{Lazarian} \& {Hoang}}{2007a}]{Lazarian_2007}
{Lazarian} A.,  {Hoang} T.,  2007a, \mn@doi [\mnras] {10.1111/j.1365-2966.2007.11817.x}, \href {https://ui.adsabs.harvard.edu/abs/2007MNRAS.378..910L} {378, 910}

\bibitem[\protect\citeauthoryear{{Lazarian} \& {Hoang}}{{Lazarian} \& {Hoang}}{2007b}]{2007MNRAS.378..910L}
{Lazarian} A.,  {Hoang} T.,  2007b, \mn@doi [\mnras] {10.1111/j.1365-2966.2007.11817.x}, \href {https://ui.adsabs.harvard.edu/abs/2007MNRAS.378..910L} {378, 910}

\bibitem[\protect\citeauthoryear{{Lee} et~al.,}{{Lee} et~al.}{2021}]{Lee_2021}
{Lee} D.,  et~al., 2021, \mn@doi [\apj] {10.3847/1538-4357/ac0cf2}, \href {https://ui.adsabs.harvard.edu/abs/2021ApJ...918...39L} {918, 39}

\bibitem[\protect\citeauthoryear{{Li}, {Fang}, {Henning}  \& {Kainulainen}}{{Li} et~al.}{2013}]{Li_2013}
{Li} H.-b.,  {Fang} M.,  {Henning} T.,   {Kainulainen} J.,  2013, \mn@doi [\mnras] {10.1093/mnras/stt1849}, \href {https://ui.adsabs.harvard.edu/abs/2013MNRAS.436.3707L} {436, 3707}

\bibitem[\protect\citeauthoryear{{Magalhães}}{{Magalhães}}{2012}]{magalhaes2012}
{Magalhães} V. d.~S.,  2012, Master dissertation, Instituto Nacional de Pesquisas Espaciais, São José dos Campos/SP, Brazil

\bibitem[\protect\citeauthoryear{{Magalhães}, {Benedetti}  \& {Roland}}{{Magalhães} et~al.}{1984}]{1984PASP...96..383M}
{Magalhães} A.~M.,  {Benedetti} E.,   {Roland} E.~H.,  1984, \mn@doi [\pasp] {10.1086/131351}, \href {https://ui.adsabs.harvard.edu/abs/1984PASP...96..383M} {96, 383}

\bibitem[\protect\citeauthoryear{{Magalhães}, {Rodrigues}, {Margoniner}, {Pereyra}  \& {Heathcote}}{{Magalhães} et~al.}{1996}]{Magalhaes1996}
{Magalhães} A.~M.,  {Rodrigues} C.~V.,  {Margoniner} V.~E.,  {Pereyra} A.,   {Heathcote} S.,  1996, in {Roberge} W.~G.,  {Whittet} D. C.~B.,  eds,  Astronomical Society of the Pacific Conference Series Vol. 97, Polarimetry of the Interstellar Medium. p.~118

\bibitem[\protect\citeauthoryear{{Marka}, {Schreyer}, {Launhardt}, {Semenov}  \& {Henning}}{{Marka} et~al.}{2012}]{Marka_2012}
{Marka} C.,  {Schreyer} K.,  {Launhardt} R.,  {Semenov} D.~A.,   {Henning} T.,  2012, \mn@doi [\aap] {10.1051/0004-6361/201014375}, \href {https://ui.adsabs.harvard.edu/abs/2012A&A...537A...4M} {537, A4}

\bibitem[\protect\citeauthoryear{{Matthews} \& {Wilson}}{{Matthews} \& {Wilson}}{2000}]{Matthews_2000}
{Matthews} B.~C.,  {Wilson} C.~D.,  2000, in American Astronomical Society Meeting Abstracts. p. 18.02

\bibitem[\protect\citeauthoryear{{Mestel}}{{Mestel}}{1966}]{Mestel1966}
{Mestel} L.,  1966, \mn@doi [\mnras] {10.1093/mnras/133.2.265}, \href {https://ui.adsabs.harvard.edu/abs/1966MNRAS.133..265M} {133, 265}

\bibitem[\protect\citeauthoryear{{Moreira}, {Yun}, {Vazquez}  \& {Torrelles}}{{Moreira} et~al.}{1997}]{Moreira_1997}
{Moreira} M.~C.,  {Yun} J.~L.,  {Vazquez} R.,   {Torrelles} J.~M.,  1997, \mn@doi [\aj] {10.1086/118350}, \href {https://ui.adsabs.harvard.edu/abs/1997AJ....113.1371M} {113, 1371}

\bibitem[\protect\citeauthoryear{{Moreira}, {Yun}, {Torrelles}, {Afonso}  \& {Santos}}{{Moreira} et~al.}{1999}]{Moreira_1999}
{Moreira} M.~C.,  {Yun} J.~L.,  {Torrelles} J.~M.,  {Afonso} J.~M.,   {Santos} C.~A.,  1999, \mn@doi [\aj] {10.1086/301018}, \href {https://ui.adsabs.harvard.edu/abs/1999AJ....118.1315M} {118, 1315}

\bibitem[\protect\citeauthoryear{{Mouschovias} \& {Ciolek}}{{Mouschovias} \& {Ciolek}}{1999}]{Mouschovias1999}
{Mouschovias} T.~C.,  {Ciolek} G.~E.,  1999, in {Lada} C.~J.,  {Kylafis} N.~D.,  eds,  NATO Advanced Study Institute (ASI) Series C Vol. 540, The Origin of Stars and Planetary Systems. p.~305

\bibitem[\protect\citeauthoryear{{Myers} \& {Basu}}{{Myers} \& {Basu}}{2021}]{Myers_2021}
{Myers} P.~C.,  {Basu} S.,  2021, \mn@doi [\apj] {10.3847/1538-4357/abf4c8}, \href {https://ui.adsabs.harvard.edu/abs/2021ApJ...917...35M} {917, 35}

\bibitem[\protect\citeauthoryear{{Ostriker}, {Stone}  \& {Gammie}}{{Ostriker} et~al.}{2001}]{Ostriker_2001}
{Ostriker} E.~C.,  {Stone} J.~M.,   {Gammie} C.~F.,  2001, \mn@doi [\apj] {10.1086/318290}, \href {https://ui.adsabs.harvard.edu/abs/2001ApJ...546..980O} {546, 980}

\bibitem[\protect\citeauthoryear{{Otrupcek}, {Hartley}  \& {Wang}}{{Otrupcek} et~al.}{2000}]{Otrupcek_2000}
{Otrupcek} R.~E.,  {Hartley} M.,   {Wang} J.~S.,  2000, \mn@doi [\pasa] {10.1071/AS00092}, \href {https://ui.adsabs.harvard.edu/abs/2000PASA...17...92O} {17, 92}

\bibitem[\protect\citeauthoryear{{Pattle} et~al.,}{{Pattle} et~al.}{2019}]{Pattle_2019}
{Pattle} K.,  et~al., 2019, \mn@doi [\apj] {10.3847/1538-4357/ab286f}, \href {https://ui.adsabs.harvard.edu/abs/2019ApJ...880...27P} {880, 27}

\bibitem[\protect\citeauthoryear{{Pattle} et~al.,}{{Pattle} et~al.}{2022}]{Pattle_2022}
{Pattle} K.,  et~al., 2022, \mn@doi [\mnras] {10.1093/mnras/stac1356}, \href {https://ui.adsabs.harvard.edu/abs/2022MNRAS.515.1026P} {515, 1026}

\bibitem[\protect\citeauthoryear{{Pattle}, {Fissel}, {Tahani}, {Liu}  \& {Ntormousi}}{{Pattle} et~al.}{2023}]{Pattle_2023}
{Pattle} K.,  {Fissel} L.,  {Tahani} M.,  {Liu} T.,   {Ntormousi} E.,  2023, in {Inutsuka} S.,  {Aikawa} Y.,  {Muto} T.,  {Tomida} K.,   {Tamura} M.,  eds,  Astronomical Society of the Pacific Conference Series Vol. 534, Protostars and Planets VII. p.~193 (\mn@eprint {arXiv} {2203.11179}), \mn@doi{10.48550/arXiv.2203.11179}

\bibitem[\protect\citeauthoryear{{Pereyra}}{{Pereyra}}{2000}]{pereyra2000}
{Pereyra} A.,  2000, PhD thesis, University of Sao Paulo, Institute for Astronomy, Geophysics, and Atmospheric Sciences

\bibitem[\protect\citeauthoryear{{Pereyra}, {Magalhaes}, {Rodrigues}  \& {Carciofi}}{{Pereyra} et~al.}{2018}]{pereyra2018}
{Pereyra} A.,  {Magalhaes} A.~M.,  {Rodrigues} C.,   {Carciofi} A.,  2018, {PCCDPACK: Polarimetry with CCD}, Astrophysics Source Code Library, record ascl:1809.002

\bibitem[\protect\citeauthoryear{{Pillai}, {Kauffmann}, {Tan}, {Goldsmith}, {Carey}  \& {Menten}}{{Pillai} et~al.}{2015}]{Pillai_2015}
{Pillai} T.,  {Kauffmann} J.,  {Tan} J.~C.,  {Goldsmith} P.~F.,  {Carey} S.~J.,   {Menten} K.~M.,  2015, \mn@doi [\apj] {10.1088/0004-637X/799/1/74}, \href {https://ui.adsabs.harvard.edu/abs/2015ApJ...799...74P} {799, 74}

\bibitem[\protect\citeauthoryear{Pillai, Clemens, Reissl  \& et al.}{Pillai et~al.}{2020}]{Pillai_2020}
Pillai T.~G.,  Clemens D.~P.,  Reissl S.,   et al. 2020, \mn@doi [Nature Astronomy] {10.1038/s41550-020-1172-6}, 4, 1195

\bibitem[\protect\citeauthoryear{{Planck Collaboration} et~al.,}{{Planck Collaboration} et~al.}{2016}]{Planck_2016}
{Planck Collaboration} et~al., 2016, \mn@doi [\aap] {10.1051/0004-6361/201525896}, \href {https://ui.adsabs.harvard.edu/abs/2016A&A...586A.138P} {586, A138}

\bibitem[\protect\citeauthoryear{{Prokopjeva}, {Sen}, {Il'in}, {Voshchinnikov}  \& {Gupta}}{{Prokopjeva} et~al.}{2014}]{Prokopjeva_2014}
{Prokopjeva} M.~S.,  {Sen} A.~K.,  {Il'in} V.~B.,  {Voshchinnikov} N.~V.,   {Gupta} R.,  2014, \mn@doi [\jqsrt] {10.1016/j.jqsrt.2014.02.017}, \href {https://ui.adsabs.harvard.edu/abs/2014JQSRT.146..410P} {146, 410}

\bibitem[\protect\citeauthoryear{Racca, Vilas-Boas  \& de~la Reza}{Racca et~al.}{2009}]{Racca_2009}
Racca G.~A.,  Vilas-Boas J. W.~S.,   de~la Reza R.,  2009, \mn@doi [The Astrophysical Journal] {10.1088/0004-637x/703/2/1444}, 703, 1444

\bibitem[\protect\citeauthoryear{{Reipurth}}{{Reipurth}}{2008}]{Reipurth_2008}
{Reipurth} B.,  2008, in {Reipurth} B.,  ed., , Vol.~5, Handbook of Star Forming Regions, Volume II.
p.~847

\bibitem[\protect\citeauthoryear{{Rodrigues}, {Magalh{\~a}es}, {Vilas-Boas}, {Racca}  \& {Pereyra}}{{Rodrigues} et~al.}{2014}]{2014IAUS..302...21R}
{Rodrigues} C.~V.,  {Magalh{\~a}es} V. d.~S.,  {Vilas-Boas} J.~W.,  {Racca} G.,   {Pereyra} A.,  2014, in {Petit} P.,  {Jardine} M.,   {Spruit} H.~C.,  eds,  IAU Symposium Vol. 302, Magnetic Fields throughout Stellar Evolution. pp 21--24 (\mn@eprint {arXiv} {1309.7599}), \mn@doi{10.1017/S1743921314001641}

\bibitem[\protect\citeauthoryear{{Sadavoy} et~al.,}{{Sadavoy} et~al.}{2018}]{Sadavoy_2018}
{Sadavoy} S.~I.,  et~al., 2018, \mn@doi [\apj] {10.3847/1538-4357/aaa080}, \href {https://ui.adsabs.harvard.edu/abs/2018ApJ...852..102S} {852, 102}

\bibitem[\protect\citeauthoryear{{Seifried}, {Walch}, {Weis}, {Reissl}, {Soler}, {Klessen}  \& {Joshi}}{{Seifried} et~al.}{2020}]{Seifried_2020}
{Seifried} D.,  {Walch} S.,  {Weis} M.,  {Reissl} S.,  {Soler} J.~D.,  {Klessen} R.~S.,   {Joshi} P.~R.,  2020, \mn@doi [\mnras] {10.1093/mnras/staa2231}, \href {https://ui.adsabs.harvard.edu/abs/2020MNRAS.497.4196S} {497, 4196}

\bibitem[\protect\citeauthoryear{{Sen}, {Gupta}, {Ramaprakash}  \& {Tandon}}{{Sen} et~al.}{2000}]{Sen_2000}
{Sen} A.~K.,  {Gupta} R.,  {Ramaprakash} A.~N.,   {Tandon} S.~N.,  2000, \mn@doi [\aaps] {10.1051/aas:2000117}, \href {https://ui.adsabs.harvard.edu/abs/2000A&AS..141..175S} {141, 175}

\bibitem[\protect\citeauthoryear{{Sen}, {Mukai}, {Gupta}  \& {Das}}{{Sen} et~al.}{2005}]{Sen_2005}
{Sen} A.~K.,  {Mukai} T.,  {Gupta} R.,   {Das} H.~S.,  2005, \mn@doi [\mnras] {10.1111/j.1365-2966.2005.09153.x}, \href {https://ui.adsabs.harvard.edu/abs/2005MNRAS.361..177S} {361, 177}

\bibitem[\protect\citeauthoryear{{Soam} et~al.,}{{Soam} et~al.}{2018}]{Soam_2018}
{Soam} A.,  et~al., 2018, \mn@doi [\apj] {10.3847/1538-4357/aac4a6}, \href {https://ui.adsabs.harvard.edu/abs/2018ApJ...861...65S} {861, 65}

\bibitem[\protect\citeauthoryear{{Soler} \& {Hennebelle}}{{Soler} \& {Hennebelle}}{2017}]{Soler_2017}
{Soler} J.~D.,  {Hennebelle} P.,  2017, \mn@doi [\aap] {10.1051/0004-6361/201731049}, \href {https://ui.adsabs.harvard.edu/abs/2017A&A...607A...2S} {607, A2}

\bibitem[\protect\citeauthoryear{{Stephens} et~al.,}{{Stephens} et~al.}{2017}]{Stephens_2017}
{Stephens} I.~W.,  et~al., 2017, \mn@doi [\apj] {10.3847/1538-4357/aa8262}, \href {https://ui.adsabs.harvard.edu/abs/2017ApJ...846...16S} {846, 16}

\bibitem[\protect\citeauthoryear{{Tassis}, {Dowell}, {Hildebrand}, {Kirby}  \& {Vaillancourt}}{{Tassis} et~al.}{2009}]{Tassis_2009}
{Tassis} K.,  {Dowell} C.~D.,  {Hildebrand} R.~H.,  {Kirby} L.,   {Vaillancourt} J.~E.,  2009, \mn@doi [\mnras] {10.1111/j.1365-2966.2009.15420.x}, \href {https://ui.adsabs.harvard.edu/abs/2009MNRAS.399.1681T} {399, 1681}

\bibitem[\protect\citeauthoryear{{Tody}}{{Tody}}{1986}]{iraf1}
{Tody} D.,  1986, in {Crawford} D.~L.,  ed.,  Society of Photo-Optical Instrumentation Engineers (SPIE) Conference Series Vol. 627, Instrumentation in astronomy VI. p.~733, \mn@doi{10.1117/12.968154}

\bibitem[\protect\citeauthoryear{{Tody}}{{Tody}}{1993}]{iraf2}
{Tody} D.,  1993, in {Hanisch} R.~J.,  {Brissenden} R.~J.~V.,   {Barnes} J.,  eds,  Astronomical Society of the Pacific Conference Series Vol. 52, Astronomical Data Analysis Software and Systems II. p.~173

\bibitem[\protect\citeauthoryear{{Tram} \& {Hoang}}{{Tram} \& {Hoang}}{2022}]{Tram_2022}
{Tram} L.~N.,  {Hoang} T.,  2022, \mn@doi [Frontiers in Astronomy and Space Sciences] {10.3389/fspas.2022.923927}, \href {https://ui.adsabs.harvard.edu/abs/2022FrASS...9.3927T} {9, 923927}

\bibitem[\protect\citeauthoryear{{Vall{\'e}e}, {Greaves}  \& {Fiege}}{{Vall{\'e}e} et~al.}{2003}]{Vallee_2003}
{Vall{\'e}e} J.~P.,  {Greaves} J.~S.,   {Fiege} J.~D.,  2003, \mn@doi [\apj] {10.1086/374309}, \href {https://ui.adsabs.harvard.edu/abs/2003ApJ...588..910V} {588, 910}

\bibitem[\protect\citeauthoryear{{Wang} \& {Chen}}{{Wang} \& {Chen}}{2019}]{Wang_2019}
{Wang} S.,  {Chen} X.,  2019, \mn@doi [\apj] {10.3847/1538-4357/ab1c61}, \href {https://ui.adsabs.harvard.edu/abs/2019ApJ...877..116W} {877, 116}

\bibitem[\protect\citeauthoryear{{Wang}, {Evans}, {Zhou}  \& {Clemens}}{{Wang} et~al.}{1995}]{Wang_1995}
{Wang} Y.,  {Evans} Neal~J. I.,  {Zhou} S.,   {Clemens} D.~P.,  1995, \mn@doi [\apj] {10.1086/176478}, \href {https://ui.adsabs.harvard.edu/abs/1995ApJ...454..217W} {454, 217}

\bibitem[\protect\citeauthoryear{{Ward-Thompson}, {Sen}, {Kirk}  \& {Nutter}}{{Ward-Thompson} et~al.}{2009}]{Ward-Thompson_2009}
{Ward-Thompson} D.,  {Sen} A.~K.,  {Kirk} J.~M.,   {Nutter} D.,  2009, \mn@doi [\mnras] {10.1111/j.1365-2966.2009.15159.x}, \href {https://ui.adsabs.harvard.edu/abs/2009MNRAS.398..394W} {398, 394}

\bibitem[\protect\citeauthoryear{{Wolf}, {Launhardt}  \& {Henning}}{{Wolf} et~al.}{2003}]{Wolf_2003}
{Wolf} S.,  {Launhardt} R.,   {Henning} T.,  2003, \mn@doi [\apj] {10.1086/375622}, \href {https://ui.adsabs.harvard.edu/abs/2003ApJ...592..233W} {592, 233}

\bibitem[\protect\citeauthoryear{{Yen} et~al.,}{{Yen} et~al.}{2020}]{Yen_2020}
{Yen} H.-W.,  et~al., 2020, \mn@doi [\apj] {10.3847/1538-4357/ab7eb3}, \href {https://ui.adsabs.harvard.edu/abs/2020ApJ...893...54Y} {893, 54}

\bibitem[\protect\citeauthoryear{{Yun} \& {Clemens}}{{Yun} \& {Clemens}}{1992}]{Yun_1992}
{Yun} J.~L.,  {Clemens} D.~P.,  1992, \mn@doi [\apjl] {10.1086/186268}, \href {https://ui.adsabs.harvard.edu/abs/1992ApJ...385L..21Y} {385, L21}

\bibitem[\protect\citeauthoryear{{Yun} \& {Clemens}}{{Yun} \& {Clemens}}{1994}]{Yun_1994}
{Yun} J.~L.,  {Clemens} D.~P.,  1994, \mn@doi [\apjs] {10.1086/191963}, \href {https://ui.adsabs.harvard.edu/abs/1994ApJS...92..145Y} {92, 145}

\bibitem[\protect\citeauthoryear{{Yun} \& {Clemens}}{{Yun} \& {Clemens}}{1995}]{Yun_1995}
{Yun} J.~L.,  {Clemens} D.~P.,  1995, \mn@doi [\aj] {10.1086/117317}, \href {https://ui.adsabs.harvard.edu/abs/1995AJ....109..742Y} {109, 742}

\bibitem[\protect\citeauthoryear{{Zielinski}, {Wolf}  \& {Brunngr{\"a}ber}}{{Zielinski} et~al.}{2021}]{Zielinski_2021}
{Zielinski} N.,  {Wolf} S.,   {Brunngr{\"a}ber} R.,  2021, \mn@doi [\aap] {10.1051/0004-6361/202039126}, \href {https://ui.adsabs.harvard.edu/abs/2021A&A...645A.125Z} {645, A125}

\makeatother
\end{thebibliography}



\appendix

\section{Gaia Overlayed Images}
\begin{figure*}
	\includegraphics[width=17.5cm]{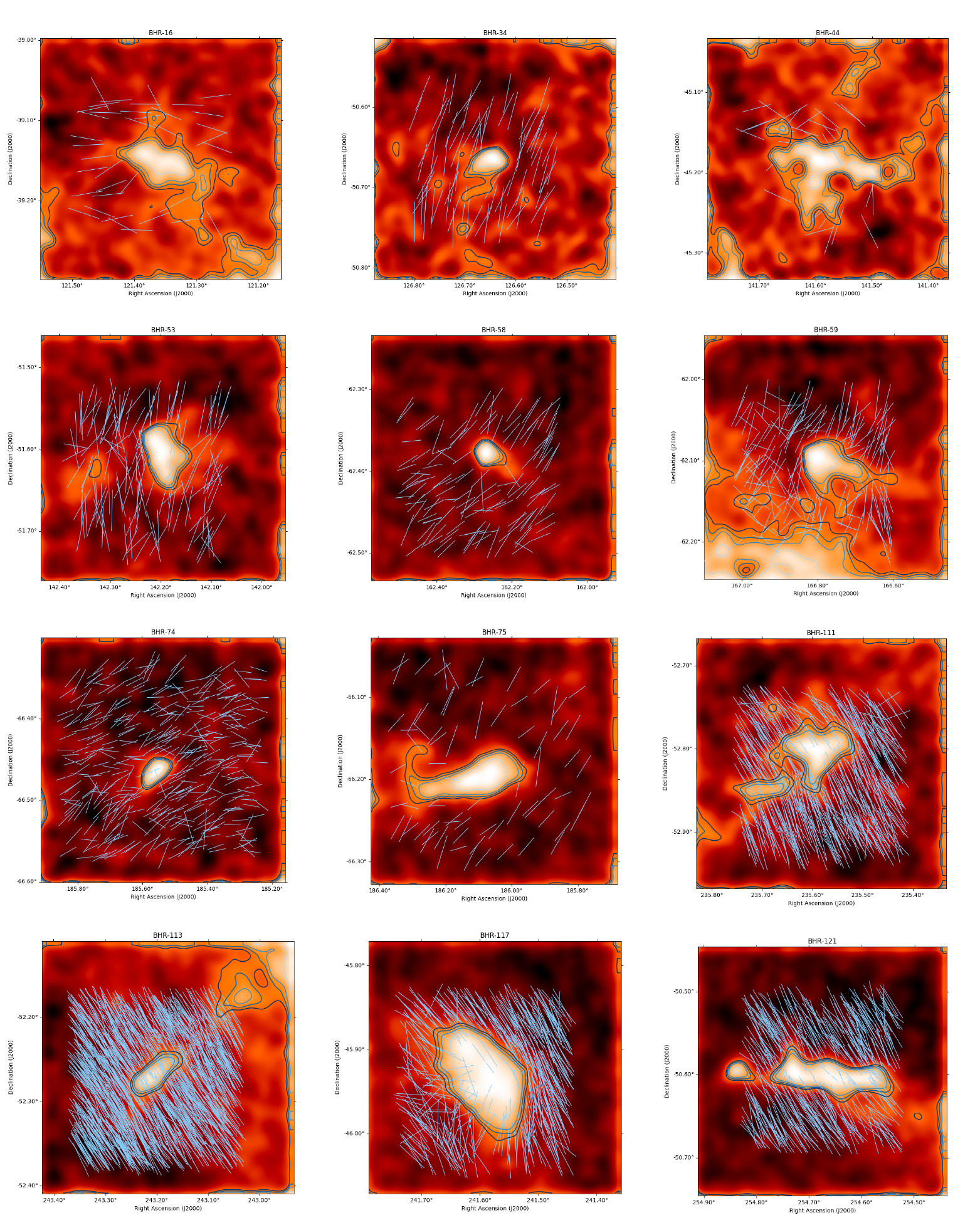}
    \caption{\textit{Gaia} Stellar Density Maps of Bok Globules}
    \label{fig:gaia-1}
\end{figure*}

\begin{figure*}
	\includegraphics[width=17.5cm]{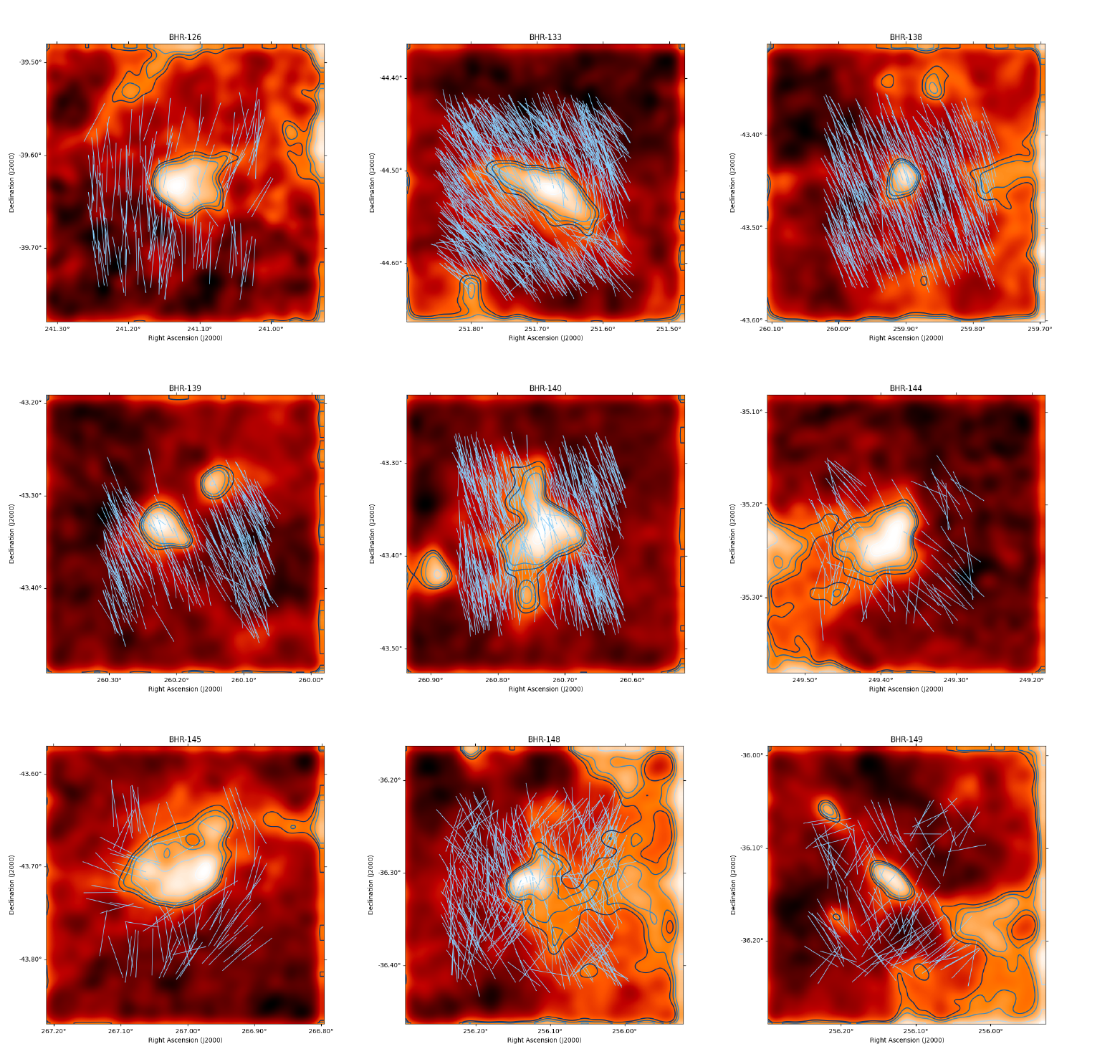}
    \caption{\textit{Gaia} Stellar Density Maps of Bok Globules}
    \label{fig:gaia-2}
\end{figure*}

The figures \ref{fig:gaia-1}~--~\ref{fig:gaia-2} show the stellar density maps obtained for each Bok globule, along with the polarization vectors of background starlight overlayed in blue.

\newpage
\newpage
\section{2MASS Overlayed Images}
\begin{figure*}
	\includegraphics[width=17.5cm]{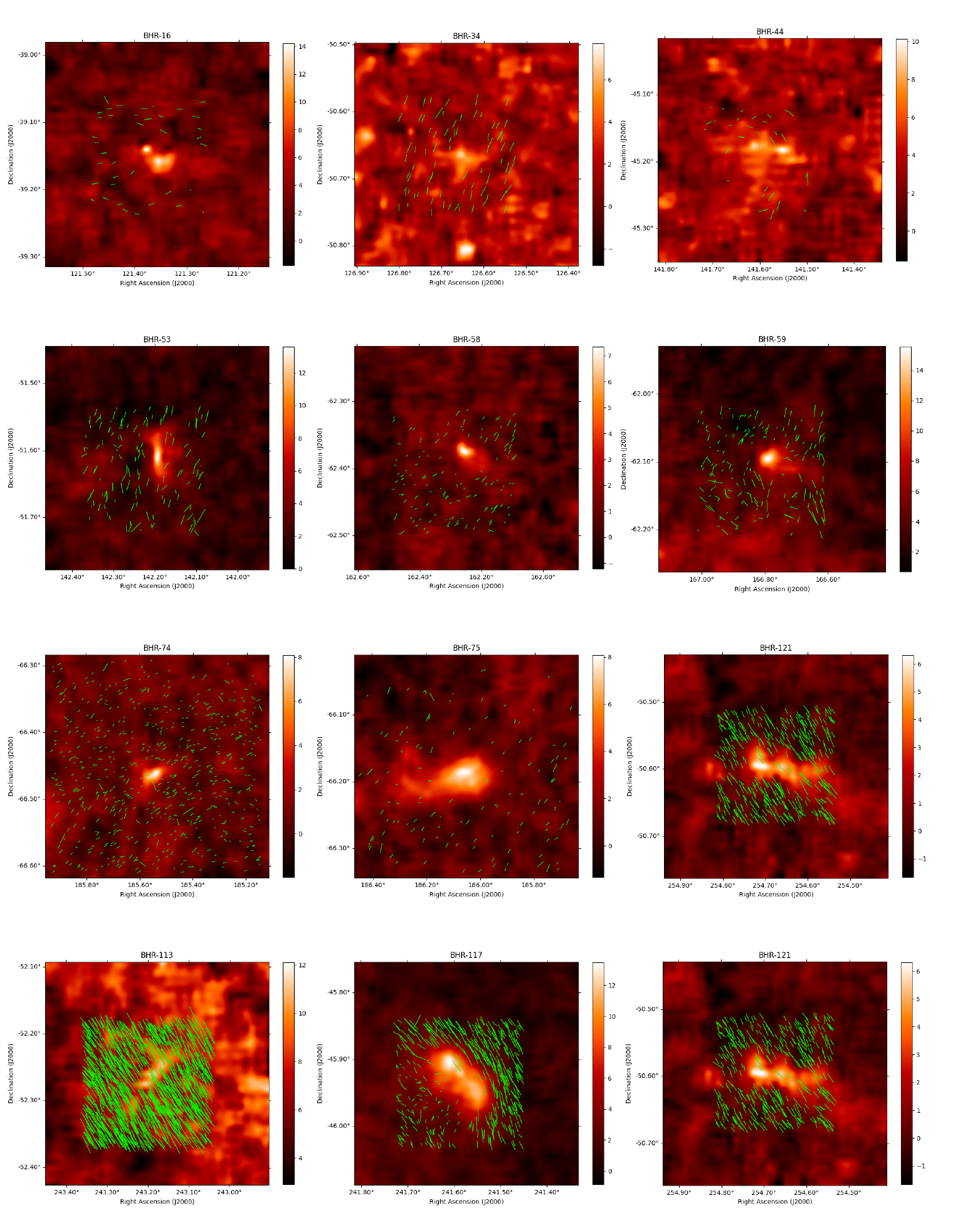}
    \caption{2MASS Extinction Maps of Bok Globules}
    \label{fig:2mass-1}
\end{figure*}

\begin{figure*}
	\includegraphics[width=17.5cm]{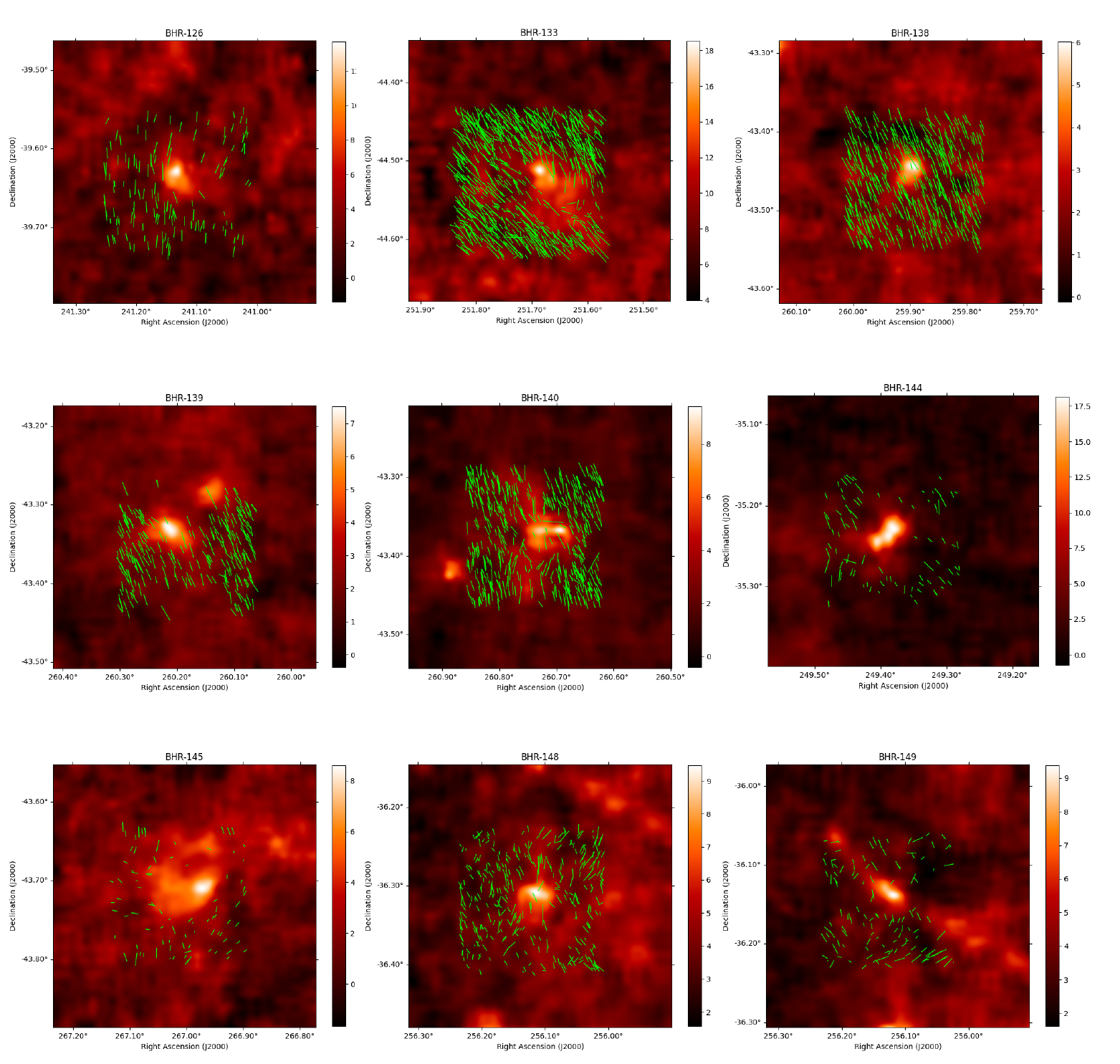}
    \caption{2MASS Extinction Maps of Bok Globules}
    \label{fig:2mass-2}
\end{figure*}

The figures \ref{fig:2mass-1}~--~\ref{fig:2mass-2} show the extinction maps obtained for each Bok globule using 2MASS reddening, along with the polarization vectors of background starlight overlayed in green. The lengths of the vectors are proportional to the polarization fraction.

\newpage
\newpage
\section{Distribution of Polarization Angles}
\begin{figure*}
	\includegraphics[width=17.5cm]{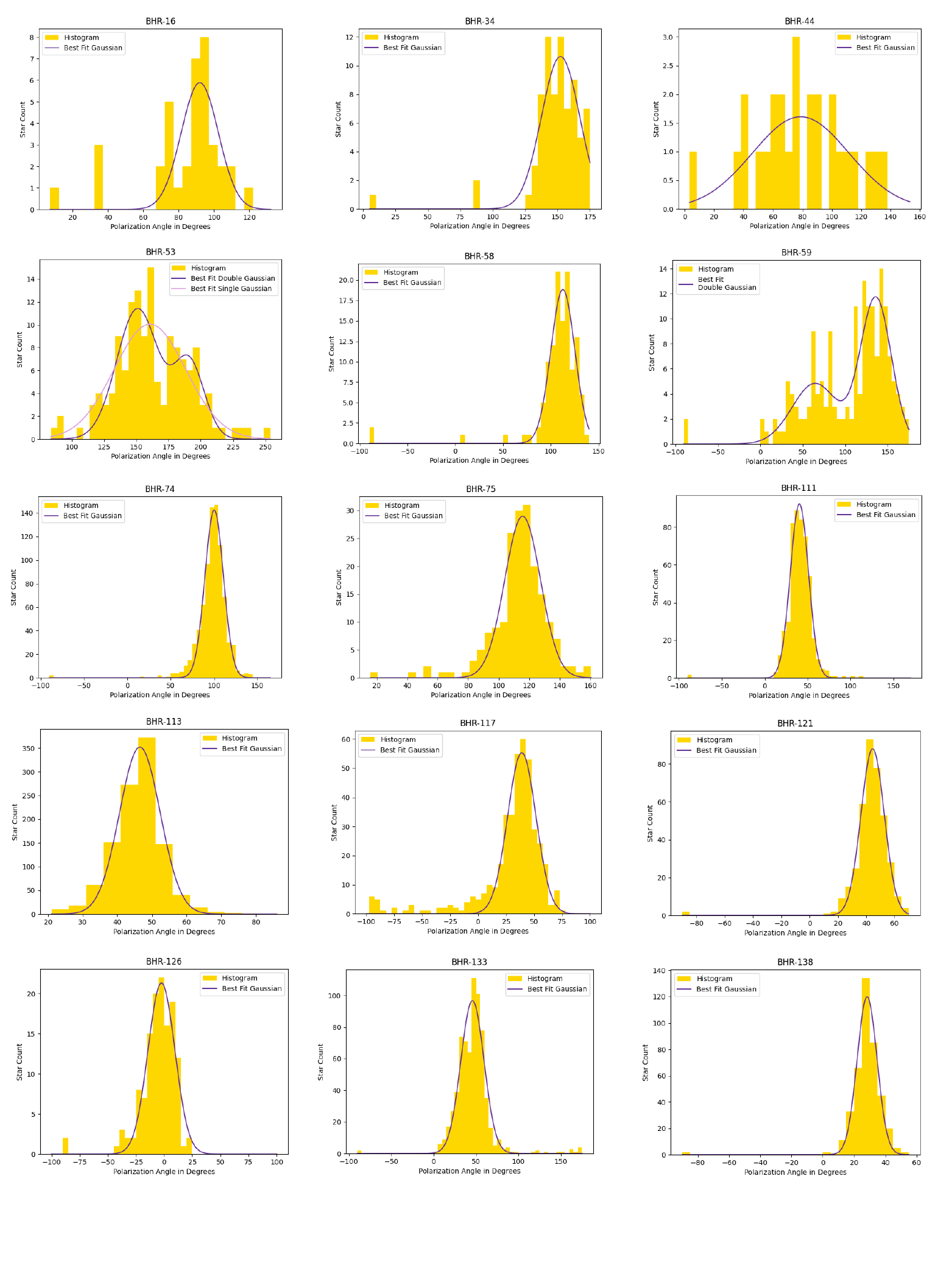}
    \caption{PA distributions with best-fitting Gaussians}
    \label{fig:pa-distribution-1}
\end{figure*}

\begin{figure*}
	\includegraphics[width=17.5cm]{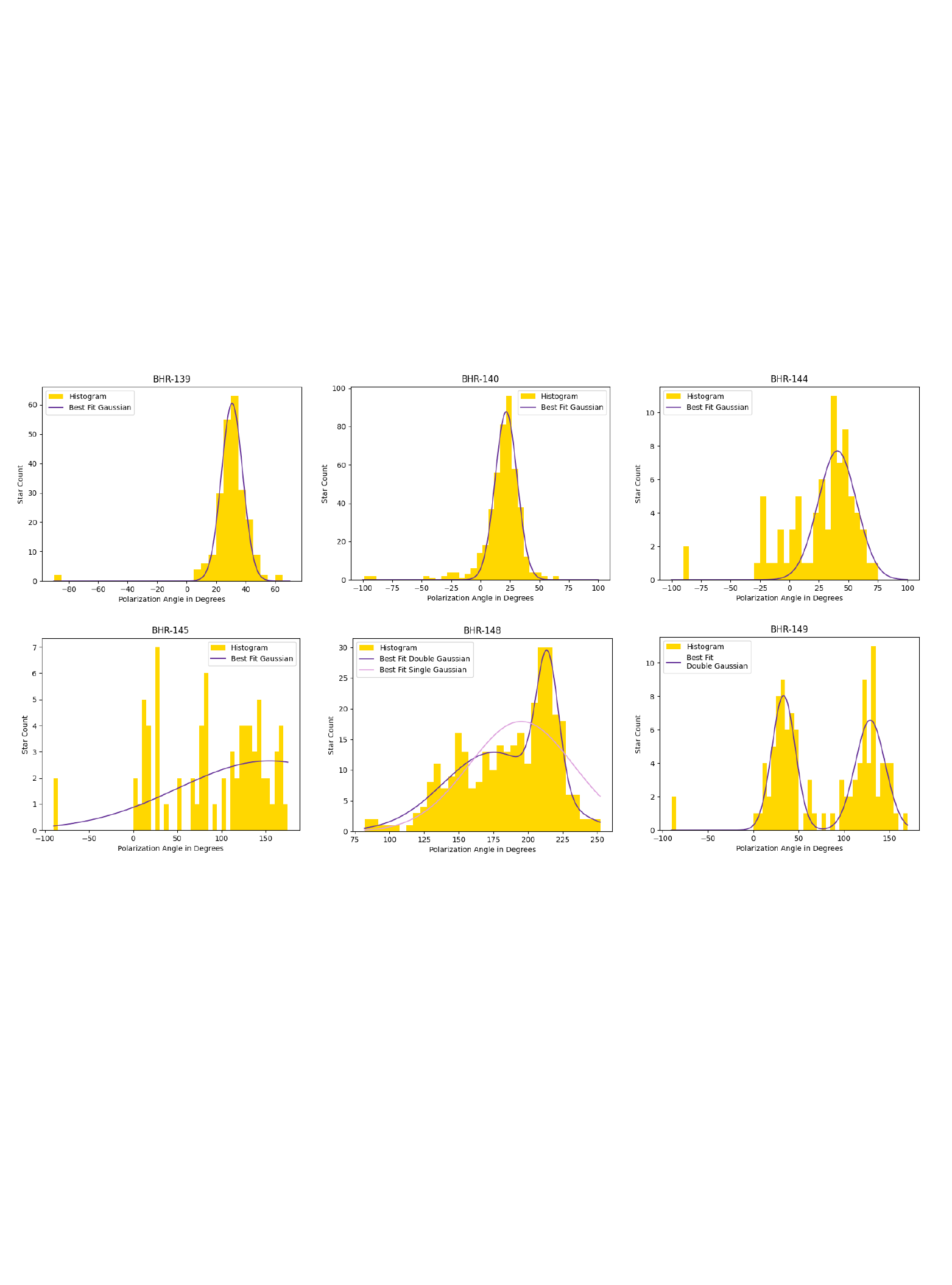}
    \caption{PA distributions with best-fitting Gaussians}
    \label{fig:pa-distribution-2}
\end{figure*}

The figures \ref{fig:pa-distribution-1}~--~\ref{fig:pa-distribution-2} show the distribution of polarization position angles around each cloud, along with the best-fit Gaussian profiles for each.

\newpage
\newpage
\section{Distribution of Polarization Fractions}
\begin{figure*}
	\includegraphics[width=17.5cm]{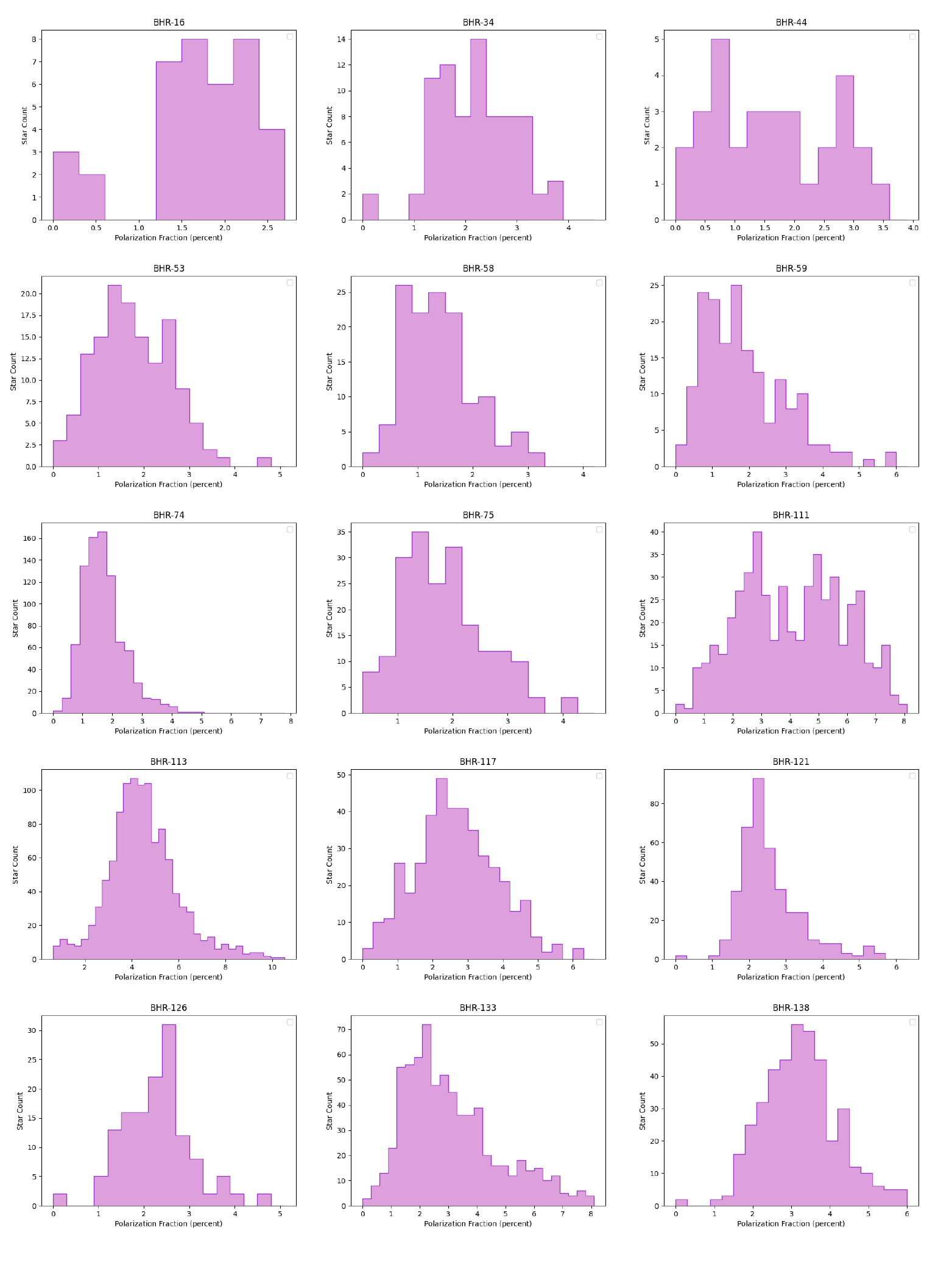}
    \caption{Polarization fraction distributions}
    \label{fig:pfracs-1}
\end{figure*}

\begin{figure*}
	\includegraphics[width=17.5cm]{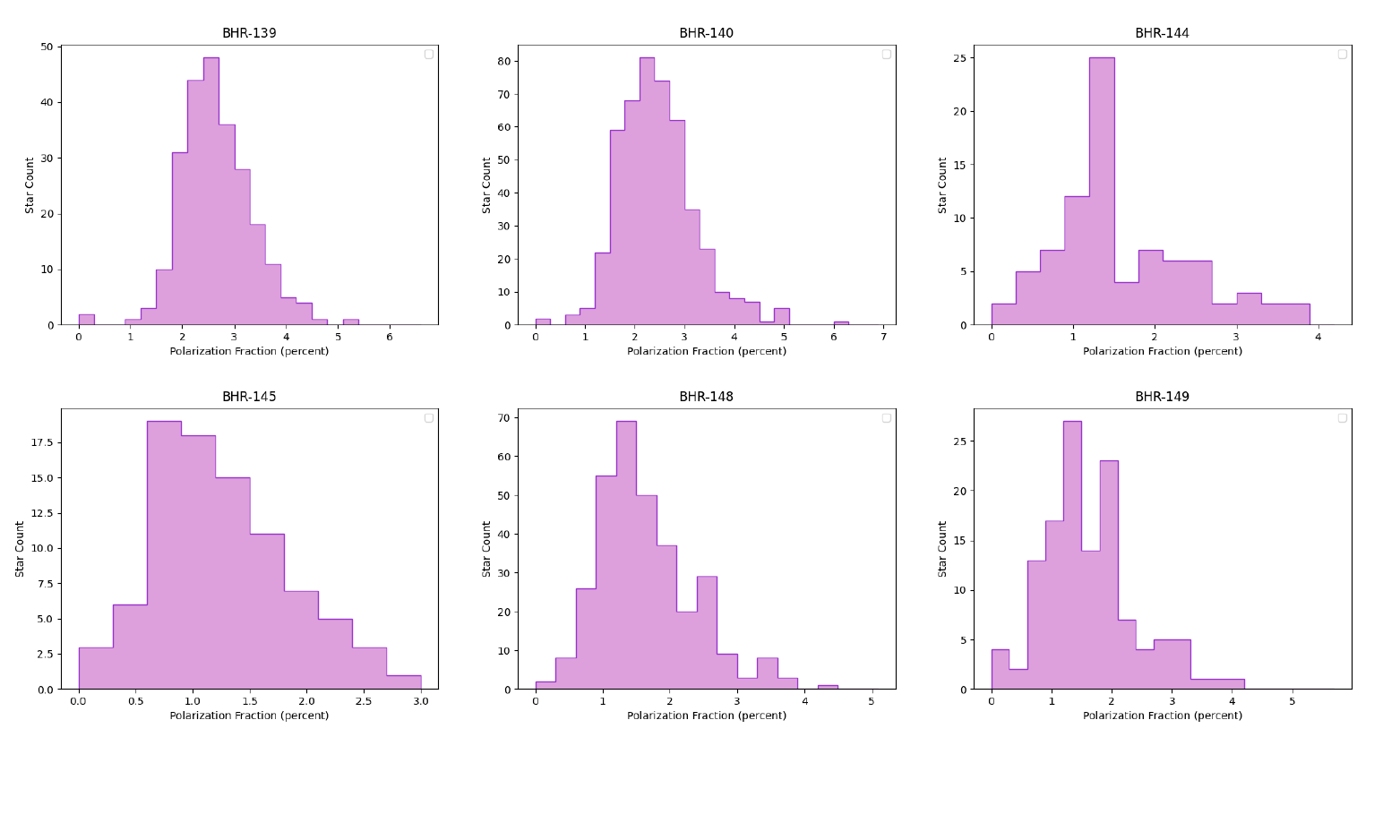}
    \caption{Polarization fraction distributions}
    \label{fig:pfracs-2}
\end{figure*}

The figures \ref{fig:pfracs-1}~--~\ref{fig:pfracs-2} show the distribution of the polarization fractions around each Bok globule.


\bsp	
\label{lastpage}
\end{document}